\documentstyle[aps,prb,eqsecnum,psfig,twocolumn,floats]{revtex}
\begin{document}
\draft
\twocolumn[\hsize\textwidth\columnwidth\hsize\csname@twocolumnfalse\endcsname
\title{Intermediate phase in the spiral antiferromagnet 
Ba$_{2}$CuGe$_{2}$O$_{7}$} 

\author{J. Chovan$^{*}$ and N. Papanicolaou \cite{NP_e-mail}}
\address{Department of Physics, University of Crete, and 
Research Center of Crete, Heraklion, Greece}
\author{S. Komineas}
\address{Physikalisches Institut, Universit\"at Bayreuth,
          D-95440, Bayreuth, Germany}

\maketitle
\date{\today}

\begin{abstract}
The magnetic compound Ba$_{2}$CuGe$_{2}$O$_{7}$ has recently been shown
to be an essentially two-dimensional spiral antiferromagnet that exhibits
an incommensurate-to-commensurate phase transition when a magnetic field
applied along the $c$-axis exceeds a certain critical value $H_c$. 
The $T=0$ dynamics is described here in terms of a continuum field theory 
in the form of a nonlinear $\sigma$ model. We are thus in a position 
to carry out  a complete calculation of the low-energy magnon spectrum 
for any strength of the applied field throughout the phase transition. 
In particular, our spin-wave analysis reveals field-induced instabilities 
at two distinct critical fields $H_1$ and $H_2$ such that $H_1<H_c<H_2$.
Hence we predict the existence of an intermediate phase whose detailed
nature is also studied to some extent in the present paper.
\\
\\
\end{abstract}

]
\section{Introduction}
\label{sec:intro}

A recent experimental investigation \cite{1,2,3,4,5} of the magnetic
properties of Ba$_{2}$CuGe$_{2}$O$_{7}$ in its low-temperature phase
($T<T_N=3.2$ K) established the occurrence of spiral antiferromagnetic order
due to a Dzyaloshinskii-Moriya (DM) anisotropy \cite{6,7}. A schematic
illustration of the spiral abstracted from experiment
may be found in Fig. 5 of Ref. 1. It was further
demonstrated that a Dzyaloshinskii-type \cite{8} commensurate-incommensurate 
(CI) phase transition is induced by a magnetic field $H$ applied along the
$c$-axis. As the field approaches a critical value $H_c\approx 2$ T, the
spiral is highly distorted while its period (pitch) grows to infinity. 
For $H>H_c$ the ground-state configuration is thought to degenerate
into a uniform spin-flop state. This phase transition
is similar to the cholesteric-nematic transition induced
by an external magnetic field in liquid crystals \cite{9,10,11}.

It is of obvious interest to describe theoretically the magnon
excitations measured by inelastic neutron scattering \cite{5},
but progress has been hindered by the great formal complexity of 
the calculation. Here we explore a new approach in which the
original discrete system is replaced by a continuum field theory.
We are thus able for the first time to carry out a complete 
calculation of the low-energy excitation spectrum for any
strength of the applied field and any direction of spin-wave
propagation. In addition, our analysis reveals the existence 
of a new intermediate phase whose properties we examine and
compare with experiment.

In Sec. II the low-energy dynamics is described in terms of a nonlinear
$\sigma$ model that is compatible with symmetry. 
In Sec. III we present a brief demonstration of the conventional
CI transition which will provide the basis for all subsequent work.
The complete field theory is first applied in Sec. IV for an analytical
calculation of the field dependence of the magnon spectrum in the high-field
commensurate phase. Interestingly, the uniform spin-flop state is shown to be
locally stable only for $H>H_2>H_c$ where the new critical field $H_2$ is
predicted to be equal to $2.9$ T. A first contact with the measured
spectrum is also made in Sec. IV.
   
The main thrust of our calculation is presented in Sec. V where the 
determination of the magnon spectrum in the low-field spiral phase is reduced
to a quasi-one-dimensional band structure problem that is solved numerically.
While an earlier calculation \cite{5} of the spectrum at $H=0$ is
confirmed, we are also in a position to analyze existing experimental
data at nonzero field and to predict the results of possible future
experiments. A byproduct of this analysis is yet another critical field
$H_1=1.7$ T $<H_c$ beyond which the flat spiral ceases to be locally stable.
Therefore, the combined results of Secs. IV and V suggest the existence
of an intermediate phase in the field region $H_1<H<H_2$ whose nature is
studied in Sec. VI where we show that a nonflat spiral becomes energetically
favorable.
The main results are summarized in the concluding Sec. VII, while discussion
of some technical issues is relegated to two Appendices.

\section{Low-energy dynamics}
\label{sec:low-energy}
The unit cell of Ba$_{2}$CuGe$_{2}$O$_{7}$ is partially illustrated in Fig. 1
where we display only the magnetic Cu sites. The lattice constants are
$a=b=8.466$ \AA	\ and  $c=5.445$ \AA. Since the Cu atoms form a perfect
square lattice within each plane, with lattice constant 
$d=a/{\sqrt{2}}\approx 6$ \AA, it is also useful to consider the
orthogonal axes $x, y$ and $z$ obtained from the original crystal axes
$a, b$ and $c$ by a $45^{\circ}$ azimuthal rotation. The complete magnetic
lattice is formally divided into two sublattices labeled by $A$ and $B$
because the major spin interaction between nearest in-plane neighbors is
antiferromagnetic. In contrast, the interaction between out-of-plane neighbors
is ferromagnetic and weak \cite{1}. Therefore, the interlayer coupling
is not crucial for our purposes and is thus ignored in the following
discussion which concentrates on the 2D spin dynamics within each layer.
\begin{figure}
\centerline{\hbox{\psfig{figure=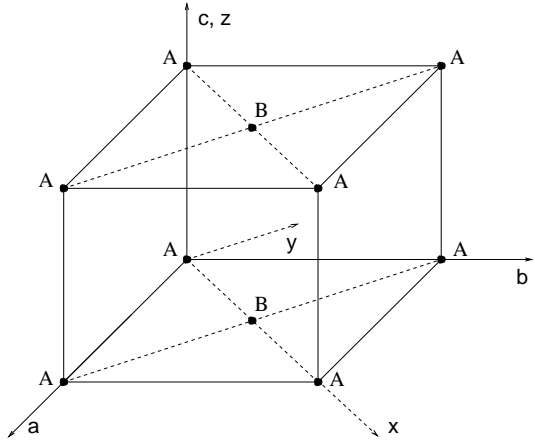,width=7.cm}}}
\vspace*{0.5cm}
\caption{Partial illustration of the unit cell of Ba$_{2}$CuGe$_{2}$O$_{7}$
 displaying only the magnetic (Cu) sites denoted by solid circles.}
\end{figure}

The space group of this crystal is $D^{3}_{2d}$ or $P\bar{4}2_{1}m$ and
imposes significant restrictions on the possible types of spin interactions.
Such symmetry constraints underlie most of the earlier work \cite{1,2,3,4,5}
but were not spelled out in sufficient detail. We have thus found it necessary
to carry out afresh a complete symmetry analysis, including both 
nearest-neighbor (nn) and next-nearest-neighbor (nnn) couplings.
For the moment, we restrict attention to nn interactions and write the 2D
spin Hamiltonian as the sum of four terms :

\begin{equation}
 \label{eq:2.1}
   W=W_{E}+W_{DM}+W_{A}+W_{Z}. 	  
\end{equation}
Here
\begin{equation}
\label{eq:2.2}
   W_{E}= \sum_{<kl>} J_{kl}({\bf S}_{k}\cdot {\bf S}_{l}) 	
\end{equation}
describes the isotropic exchange over nn in-plane bonds, denoted by
$<kl>$ , with $J_{kl}=J$ for all such bonds. Similarly,         
\begin{equation}
\label{eq:2.3}
 W_{DM}=\sum_{<kl>}{\bf D}_{kl}\cdot({\bf S}_{k}\times{\bf S}_{l})	
\end{equation}
stands for antisymmetric DM anisotropy where the vectors ${\bf D}_{kl}$
assume four distinct values:
\begin{eqnarray}
 \label{eq:2.4}
       {\bf D}_{I} &=& 
       D{\bf e}_{2} + D^{\prime}{\bf e}_{3},
            \hspace*{0.7cm}{\bf D}_{II} = D{\bf e}_{1} + D^{\prime}{\bf e}_{3},
        \nonumber
        \\    
{\bf D}_{III} &=& 
       D{\bf e}_{2} - D^{\prime}{\bf e}_{3},
            \hspace*{0.7cm}{\bf D}_{IV} = D{\bf e}_{1} - D^{\prime}{\bf e}_{3},
\end{eqnarray}
which are distributed over the 2D lattice as shown in Fig. 2 where nn bonds
are accordingly labeled by I, II, III or IV.
\begin{figure}
\centerline{\hbox{\psfig{figure=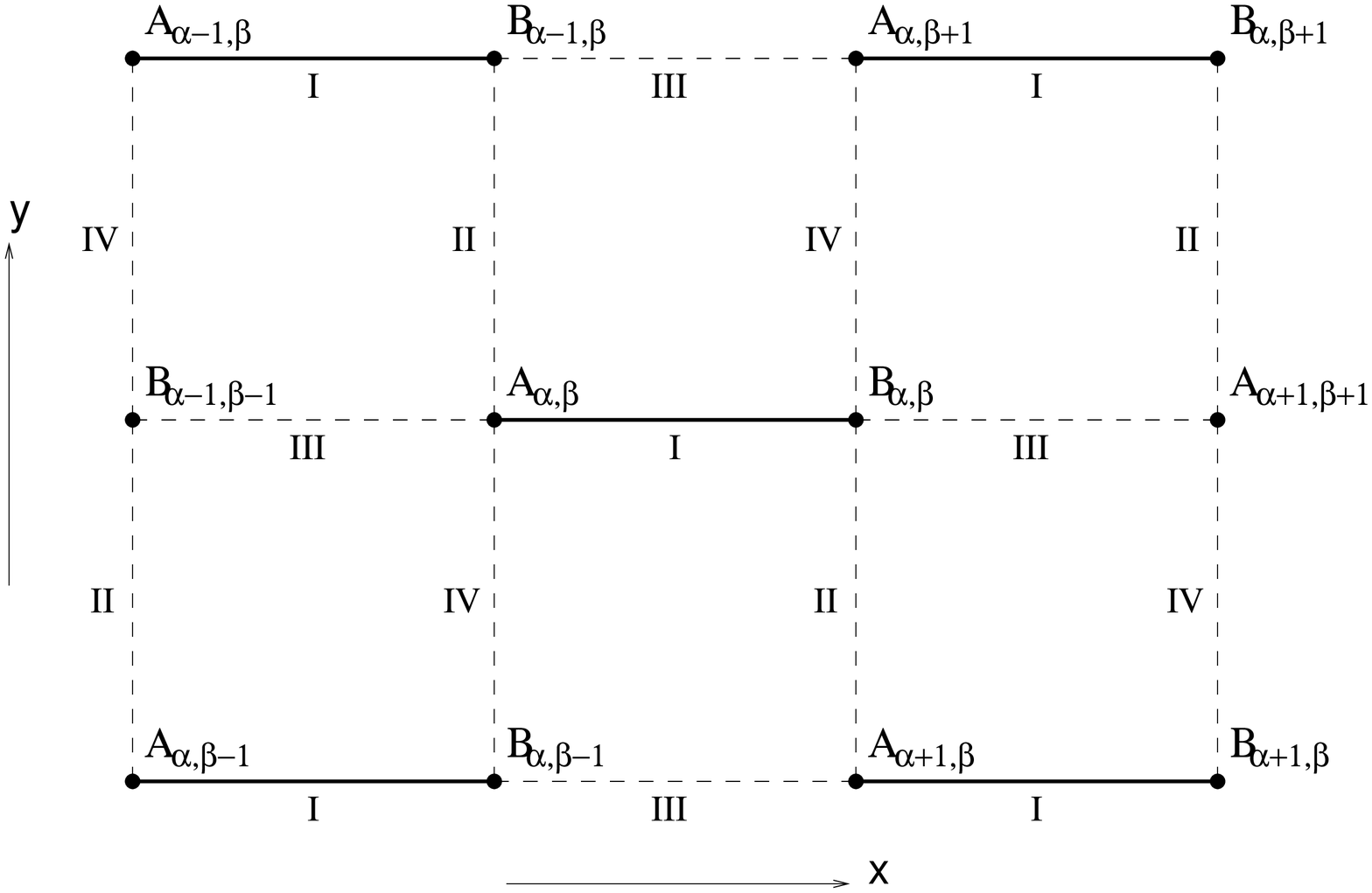,width=7.cm}}}
\vspace*{0.5cm}
\caption{ Illustration of the dimerization process on a finite portion
 of the 2D lattice cut along the axes $x$ and $y$. The indices $\alpha$
 and $\beta$ advance along the crystal axes $a$ and $b$ not shown in
 this figure. The meaning of the Roman labels on bonds connecting nn sites
 is explained in the text. }
\end{figure}
Here $D$ and $D^{\prime}$ are two
independent scalar constants, while ${\bf e}_{1}, {\bf e}_{2}$ and 
${\bf e}_{3}$ are unit vectors along the $x, y$ and $z$ axes of Fig. 1.
It should be noted that the $z$-components of the DM vectors alternate
in sign on opposite bonds, a feature that could lead to weak ferromagnetism.
No such alternation occurs for the in-plane components of the DM vectors
(\ref{eq:2.4}) which are responsible for the observed spiral magnetic order
or helimagnetism.

The third term in Eq. (\ref{eq:2.1}) contains all ``symmetric'' anisotropies.
Since single-ion anisotropy is not possible in this spin $s=\frac{1}{2}$
system, the most general form of $W_{A}$ is
\begin{equation}
 \label{eq:2.5} 
  W_{A} = \frac{1}{2}\sum_{<kl>}\sum_{i,j}G^{ij}_{kl}
         \left( S^{i}_{k}S^{j}_{l} + S^{j}_{k}S^{i}_{l} \right) ,
\end{equation}
where the indices $i$ and $j$ are summed over three values corresponding 
to the Cartesian components of the spin vectors along the axes $x,y$ and
$z$. Accordingly, $G_{kl}=(G^{ij}_{kl})$ are $3\times 3$ symmetric matrices,
one for each bond $<kl>$. Again, there exist four distinct such matrices:
\vspace{0.5cm}
{\setlength\arraycolsep{2pt}
\begin{eqnarray}
\label{eq:2.6}
G_{I}=  
\left(\begin{array}{ccc}
\hspace*{0.1cm} K_{1}& 0      &\hspace*{-0.1cm} 0       \\
 0      &\hspace*{0.1cm} K_{2}&\hspace*{0.1cm} K_{4} \\
 0      &\hspace*{0.1cm} K_{4}&\hspace*{0.1cm} K_{3}  \end{array}\right),
\quad &&
G_{II}=
\left(\begin{array}{ccc}
\hspace*{0.1cm} K_{2}& 0      &\hspace*{0.1cm} K_{4}   \\
 0      &\hspace*{0.1cm} K_{1}&\hspace*{-0.1cm} 0         \\
\hspace*{0.1cm} K_{4}& 0      &\hspace*{0.1cm} K_{3}  \end{array}\right),
      \nonumber
      \\
G_{III}=
\left(\begin{array}{ccc}
 K_{1}&\hspace*{0.1cm} 0      & 0       \\
 0      &\hspace*{0.1cm}K_{2}&\hspace*{-0.1cm}-K_{4} \\
 0      &\hspace*{-0.2cm}-K_{4}&\hspace*{0.2cm}K_{3}  \end{array}\right),
\quad   &&
G_{IV}=
\left(\begin{array}{ccc}
\hspace*{0.1cm} K_{2}& 0      &\hspace*{-0.2cm}-K_{4}   \\
\hspace*{0.1cm} 0      &\hspace*{0.1cm}K_{1}&\hspace*{-0.1cm} 0         \\
\hspace*{-0.1cm}-K_{4}& 0    &\hspace*{0.1cm} K_{3} \end{array}\right),
\nonumber \\
\end{eqnarray}}
\hspace*{-8.5pt}
which are all expressed in terms of the four scalar parameters
$K_{1}, K_{2}, K_{3}$ and $K_{4}$. The latter may be further restricted
by the trace condition $K_{1}+K_{2}+K_{3}=0$ because the isotropic
component of the exchange interaction is already accounted for by 
Eq. (\ref{eq:2.2}). Finally,
\begin{equation}
 \label{eq:2.7}
  W_{Z}= -\sum_{l}(g {\mu}_{B}{\bf H}\cdot{\bf S}_{l})
\end{equation}
describes the usual Zeeman interaction with an external field ${\bf H}$.

The discrete Hamiltonian could be employed to analyze this system
by standard spin-wave techniques, but the calculational burden is rather
significant and has so far prevented a complete determination of the magnon
spectrum \cite{5}. Nevertheless, the relevant low-energy dynamics can be
efficiently calculated in terms of a continuum field theory which provides
a reasonable approximation for Ba$_{2}$CuGe$_{2}$O$_{7}$ because the period of
the observed spiral is equal to about 37 lattice constants along the
$x$-direction. A similar approach is often invoked in the related
subject of weak ferromagnetism \cite{12,13} and can be implemented by a
straightforward step-by-step procedure starting from the original
discrete Hamiltonian \cite{14,15}.

The first step is to group spins into dimers as shown in Fig. 2. Each dimer
contains a pair of spins denoted by ${\bf A}$ and ${\bf B}$ and labeled
by a common set of sublattice indices ${\alpha}$ and ${\beta}$ that
advance along the crystal axes $a$ and $b$.  A more convenient set of
variables is given by the ``magnetization'' ${\bf m}$ and the 
``staggered magnetization'' ${\bf n}$ which are defined as
 \begin{equation}
 \label{eq:2.8}
  {\bf m}= \frac{1}{2s}({\bf A} + {\bf B}) ,\hspace*{1cm}
  {\bf n}= \frac{1}{2s}({\bf A} - {\bf B}) ,
\end{equation}
and satisfy the classical constraints ${\bf m}\cdot{\bf n}=0$ and
${\bf m}^{2} + {\bf n}^{2} = 1$. We also introduce space-time variables
according to 
\begin{equation}
 \label{eq:2.9}
     \eta =\sqrt{2}\varepsilon\alpha ,\hspace*{1cm}
     \xi = \sqrt{2}\varepsilon\beta  ,\hspace*{1cm}
     \tau = 2s\sqrt{2}\varepsilon Jt,
\end{equation}
where ${\varepsilon}$ is a dimensionless scale whose significance will become
apparent as the discussion progresses. The final result will be stated in 
terms of the coordinates
\begin{equation}
 \label{eq:2.10}
        x=\frac{{\xi} + {\eta}}{\sqrt{2}} ,\hspace*{1cm}
        y=\frac{{\xi} - {\eta}}{\sqrt{2}} ,
\end{equation}
along the $x$ and $y$ axes of Fig. 1. One should keep in mind that
actual distances are given by ${xd}/{\varepsilon}$ and
${yd}/{\varepsilon}$ where $d={a}/\sqrt{2}$ is the lattice
constant of the square lattice formed by the Cu atoms. Finally, we 
introduce rescaled anisotropy constants and magnetic field as
\begin{eqnarray}
 \label{eq:2.11}
    & \displaystyle{
    \lambda = \frac{D}{\varepsilon J} ,\hspace*{1cm}
   {\lambda}^{\prime} = \frac{\sqrt{2}D^{\prime}}{\varepsilon J} ,\hspace*{1cm}
    {\bf h} = \frac{g{\mu}_{B}{\bf H}} {2s\sqrt{2}\varepsilon J} 
}& ,
     \nonumber
     \\
     \nonumber
     \\
   & \displaystyle{
    \kappa_{0} = \frac{2}{\varepsilon^{2} J}\left(K_{1} 
  + K_{2} -2 K_{3} \right),
}&     
\end{eqnarray}
where we display only those combinations of constants that survive in the
effective low-energy dynamics. In particular, the constant $K_{4}$ does not
appear to leading order. The further notational abbreviations

\begin{equation}
\label{eq:2.12}
{\kappa}={\kappa}_{0} -{\lambda}^{2} + {{\lambda}^{\prime}}^{2} ,\hspace*{1cm}
{\bf d}_{z} = {\lambda}^{\prime} {\bf e}_{3}
\end{equation}
will prove convenient in all subsequent calculations.

Now, a consistent low-energy expansion is obtained by treating $\bf {m}$ as
a quantity of order $\varepsilon$ while $\bf {n}$ is of order unity.
To leading order, the classical constraints reduce to

\begin{equation} 
 \label{eq:2.13}
  {\bf m}\cdot{\bf n} = 0 ,\hspace*{1cm}
  {\bf n}^{2} = 1 ,
\end{equation}
$\bf {m}$ is expressed entirely in terms of $\bf {n}$ by

\begin{equation}
\label{eq:2.14}  
  {\bf m} = \frac {\varepsilon}{2\sqrt{2}}\left[{\bf n}\times
            \left(\dot{\bf n} + {\bf d}_{z} 
            - {\bf n}\times{\bf h}\right)\right]
            - \frac{\varepsilon}{2}{\partial}_{1}{\bf n} ,
\end{equation}
and the $T=0$ dynamics of the staggered magnetization $\bf {n}$ is governed
by the Lagrangian density ${\cal L}={\cal L}_{0} - V$ where
 
\begin{eqnarray}
\label{eq:2.15}
{\cal L}_{0}&=& \frac{1}{2}{\dot{\bf n}}^{2}  
                        + {\bf h}\cdot({\bf n}\times{\dot{\bf n}}) ,
	\nonumber
	\\
     V &=& {\frac{1}{2}} ({\partial}_{1}{\bf n} 
             - {\lambda} {\bf e}_{2}\times{\bf n})^{2}
             + \frac{1}{2}({\partial}_{2}{\bf n} 
             - {\lambda}{\bf e}_{1}\times{\bf n})^{2}
         \nonumber
	\\
        & &   
        +\frac{1}{2}{\kappa} n^{2}_{3} + \frac{1}{2}({\bf n}\cdot{\bf h})^{2} 
        +({\bf h}\times {\bf d}_{z})\cdot{\bf n}.
\end{eqnarray}
The dot denotes differentiation with respect to the time variable
$\tau$, ${\partial}_{1}$ and ${\partial}_{2}$ are partial derivatives with
respect to $x$ and $y$, and ($n_{1},n_{2},n_{3}$) are the Cartesian
components of $\bf {n}$ along the axes $xyz$ of Fig. 1. Consistency requires
that all physical predictions derived from Eqs. (\ref{eq:2.14}) and
(\ref{eq:2.15}) must be independent of the specific choice of the scale
parameter $\varepsilon$. This fact will be explicitly demonstrated or
used to advantage in the continuation of the paper.

We have further examined possible modifications of the 
low-energy dynamics due to nnn spin interactions along the diagonals
of the Cu plaquettes. Our symmetry analysis revealed that both
antisymmetric (DM) and symmetric anisotropies are present over nnn bonds
and introduce a new set of parameters. Nevertheless, in the continuum
limit, all new parameters merge with those already present in the
Lagrangian (\ref{eq:2.15}). The implied remarkable rigidity of the effective
low-energy spin dynamics is obviously due to the special crystal structure
of Ba$_{2}$CuGe$_{2}$O$_{7}$.

In the remainder of this section we make contact with the static energy
functional derived by Zheludev et al. \cite{5}, restricted to $T=0$,
which appears to differ in some respects from the potential $V$ of
Eq. (\ref{eq:2.15}). First, we note that we have omitted from the potential
some additive field-dependent constants which play no role except to relate
the energy to the magnetization. The latter will be obtained in Sec. III by
a direct application of Eq. (\ref{eq:2.14}).
A more interesting point concerns the special choice of exchange anisotropy
made in Ref. 5, which was suggested by the work of Kaplan \cite{16},
Shekhtman, Entin-Wohlman and Aharony \cite{17}, and is referred to as 
the KSEA anisotropy. If the original perturbative derivation of the 
antisymmetric DM interaction \cite{7} is carried to second order \cite{17},
a symmetric anisotropy results that is described by a special case
of the matrices (\ref{eq:2.6}) with
 
\begin{eqnarray}
\label{eq:2.16}
       &&      
  K_{1} = 0 , \hspace{1.4cm}
  K_{2} = \frac{D^{2}}{2J} , 
       \nonumber
       \\
       &&
       \nonumber
       \\
       &&
  K_{3} = \frac{D^{\prime2}}{2J} , \hspace{0.7cm}
  K_{4} = \frac{DD^{\prime}}{2J} ,
\end{eqnarray}
in addition to a simple renormalization of the exchange constant $J$.
The parameter ${\kappa}_{0}$ of Eq. (\ref{eq:2.11}) is then given by
${\kappa}_{0}= {\lambda}^{2}-{\lambda}^{\prime2}$ and the parameter
$\kappa$ of Eq. (\ref{eq:2.12}) vanishes. Since a nonzero $\kappa$ is
allowed by symmetry, we shall keep it throughout our theoretical development.
However, our numerical demonstrations will also be restricted to the
KSEA limit ($\kappa=0$).

Finally, the term $({\bf h}\times {\bf d}_{z})\cdot{\bf n}$ in the
potential $V$ of Eq. (\ref{eq:2.15}) is absent from the energy functional
of Zheludev et al. \cite{5}. A contribution of that nature is present
in the early work of Andreev and Marchenko \cite{12} and plays a significant
role in various aspects of weak ferromagnetism \cite{15}. This term
vanishes when the field
is applied along the $c$-axis (${\bf h}\times {\bf d}_{z}=0$) and thus
does not affect the analysis of the CI transition.
However, such a term is important in the case of an in-plane magnetic field
which is also of experimental interest \cite{3} and
is briefly discussed in the concluding paragraph of Sec. III.

\section{Ground state}
\label{sec:ground_state}

An important first step in the calculation of the $T=0$ dynamics is
the search for the classical spin configuration that minimizes the
static energy

\begin{equation}
\label{eq:3.1}
  W = \int {V\,dx dy} ,
\end{equation}
where $V$ is the potential of Eq. (\ref{eq:2.15}). For a field applied
along the $c$-axis, ${\bf h}=(0,0,h)$, the potential is given by

\begin{eqnarray}
 \label{eq:3.2}
   V  &=& \frac{1}{2}\left[ \left({\partial}_{1}{\bf n}\right)^{2} 
        + \left({\partial}_{2}{\bf n}\right)^{2}
          +{\gamma}^{2}n_{3}^{2} + {\lambda}^{2}\ \right]  
          \nonumber
          \\
     & & - \lambda \left[\left({\partial}_{1}n_{1} 
         - {\partial}_{2}n_{2}\right)n_{3}
        -\left(n_{1}{\partial}_{1} - n_{2}{\partial}_{2}\right)n_{3}\ \right] ,
\end{eqnarray}
which depends only on the parameter $\lambda$ that measures the strength
of the in-plane component of the DM anisotropy, and the combination 
of parameters 

\begin{equation}
\label{eq:3.3}
    {\gamma}^{2}= {\kappa} + {\lambda}^{2} + h^{2}
\end{equation}
that includes the external field $h$. A notable feature of the potential
(\ref{eq:3.2}) is its invariance under the simultaneous transformations

\begin{eqnarray}
\label{eq:3.4}
  &x+iy \rightarrow ( x + iy)\ e^{i {\psi}_{0}},&
   \nonumber
   \\
  &n_{1}+in_{2} \rightarrow (n_{1} + in_{2})\ e^{- i {\psi}_{0}} .&
\end{eqnarray}
This is a peculiar realization of $U(1)$ symmetry in that the usual
2D rotation of spatial coordinates with an angle ${\psi}_{0}$ is followed
by an azimuthal rotation of the staggered magnetization with an angle
$-{\psi}_{0}$.

The minimization problem was extensively studied in the earlier work
\cite{1,2,3,4,5}. Here we briefly describe a slightly simplified version
of the obtained solution in order to establish convenient notation for
our subsequent dynamical calculations.  If we invoke the usual spherical
parametrization of the unit vector ${\bf n}$ defined from

\begin{equation}
 \label{eq:3.5}
   {n}_{1}+i{n}_{2}=\sin{\Theta}\, e^{i \Phi} ,\hspace*{1cm}
   {{n}_{3}}=\cos \Theta ,
\end{equation}
the minimum of the energy is sought after in the form of the one-dimensional
(1D) Ansatz

\begin{equation}
\label{eq:3.6}
    {\Theta} = \theta(x) , \hspace*{1cm}
    {\Phi} = 0 ,
\end{equation}
which assumes that the staggered magnetization is confined in the $xz$-plane
and depends only on the spatial coordinate $x$, modulo a $U(1)$ transformation
given by Eq. (\ref{eq:3.4}). The potential (\ref{eq:3.2}) then simplifies to

\begin{equation}
 \label{eq:3.7}
V = \frac{1}{2}\left[\left({\theta}^{\prime} - \lambda \right)^{2}
        +{\gamma}^{2}{\cos}^{2}\theta \ \right],
\end{equation}
where the prime denotes differentiation with respect to $x$, and stationary 
points of the energy (\ref{eq:3.1}) satisfy the ordinary differential equation
$ {\theta}^{\prime\prime}+{\gamma}^{2}{\cos}{\theta}\ {\sin}{\theta}=0$ whose
distinct feature is that it does not depend on $\lambda$. A first integral
of this equation is given by 
${\theta}^{\prime 2} - {\gamma}^{2}{\cos}^{2}{\theta} = C = {\delta}^{2}$,
where we anticipate the fact that the minimum of the energy is achieved at
positive integration constant $C$. Thus the desired solution 
$\Theta=\theta(x)$ is given by the implicit equation                           
\begin{equation}
\label{eq:3.8}
   x=\int^{\theta}_{0} \frac{d\vartheta}{\sqrt{\delta^{2}
             + \gamma^{2}{\cos}^{2}\vartheta}} ,
\end{equation}
and is a monotonically increasing function of $x$. The corresponding
spin structure repeats itself when $\theta$ is changed by an amount $2\pi$ ; 
i.e., when $x$ advances by a distance           
\begin{equation}
\label{eq:3.9}
          L =4\int^{\frac{\pi}{2}}_{0}\frac{d\theta}{\sqrt{\delta^{2}
             + \gamma^{2}{\cos}^{2}\theta}} ,
\end{equation}
which will be called the period of the spiral. The free parameter $\delta$ is 
determined by the requirement that the average energy density 
$w=\frac{1}{L}\int^{L}_{0}Vdx$ is a minimum, where $V$ is the 
potential (\ref{eq:3.7}) calculated for the specific configuration 
(\ref{eq:3.8}). A direct computation shows that ${\delta}$ must 
satisfy the algebraic equation
\begin{equation}
\label{eq:3.10}
 {\frac {2}{\pi}}\int^{\frac{\pi}{2}}_{0} d{\theta}{\sqrt{\delta^{2}
             + \gamma^{2}{\cos}^{2}\theta}} = \lambda ,
\end{equation}
and the corresponding energy density is
\begin{equation}
\label{eq:3.11}
  w= {\frac {1}{2}}({\lambda}^{2} - {\delta}^{2}).
\end{equation}
The configuration described above will be referred to as the flat
spiral because the staggered magnetization is confined in the $xz$-plane.

It is clear that the root $\delta$ of Eq. (\ref{eq:3.10}) decreases with
increasing $\gamma$. In fact, $\delta$ vanishes at a critical value of
$\gamma$ which is easily calculated by setting $\delta = 0$ in Eq. 
(\ref{eq:3.10}) to obtain $\gamma = \gamma_{c}= \lambda {\pi}/2$.
In view of Eq. (\ref{eq:3.3}), the corresponding critical field is given by
\begin{equation}
\label{eq:3.12}
h_{c}=\left[\left( \frac {\pi}{4}^{2}-1 \right) {\lambda}^{2} 
    - \kappa \ \right]^{1/2} ,
\end{equation}
and a spiral state is possible only for $ h < h_{c}$. At the critical point, 
the energy density (\ref{eq:3.11}) becomes $w = {\lambda}^{2}/2$ and is equal
to the energy of the uniform spin-flop state $ {\bf n} = (1,0,0)$. The latter
is a stationary point of the energy functional for any strength of the applied
field and is thought to be the absolute minimum for $h > h_{c}$. The actual
stability of the spin-flop state for $h > h_{c}$, and of the spiral state
for $h < h_{c}$, will be addressed more carefully in Secs. IV and V.

Next we calculate the $T=0$ magnetization ${\bf m}=(m_{1},m_{2},m_{3})$ which
can be obtained from Eq. (\ref{eq:2.14}) applied for the static configuration
${\bf n}=({\sin} {\theta},0,{\cos} {\theta})$ and averaged over the period
$L$ of the spiral. The only term that survives in the average is
\begin{equation}
\label{eq:3.13}
 m_{3} =\frac{\varepsilon h}{2 \sqrt {2}}
        \frac{1}{L}          
\int^{ L}_{0} (1 - {\cos}^{2}{\theta})\ dx ,
\end{equation}
and can be expressed in terms of quantities already considered, namely
\begin{equation}
\label{eq:3.14}
 m_{3}=\frac{\varepsilon h}{2\sqrt{2}{\gamma}^{2}}  
       \left({\gamma}^{2} + {\delta}^{2} - \frac {2 \pi \lambda}{L}\right), 
           \hspace*{1cm} h < h_{c}.
\end{equation}
For $h > h_{c}$, the spin-flop state ${\bf n}=(1,0,0)$ is inserted in Eq.
(\ref{eq:2.14}) to yield after a trivial computation

\begin{equation}
\label{eq:3.15}
 m_{3} = \frac{{\varepsilon}h}{2{\sqrt 2}}, \hspace*{1cm} h > h_{c} ,
\end{equation}
while $m_{1}=0$ and $m_{2}= - {\varepsilon}{\lambda}^{\prime}/2{\sqrt 2}$.
The latter formula is the only place where the oscillating component 
of the DM anisotropy appears and produces a field-independent weak 
ferromagnetic moment along the $y$-axis.

In order to make definite quantitative predictions we use as input \cite{5}
the spin value $s=1/2$, an exchange constant $J=0.96$ meV, and a gyromagnetic
ratio $g=g_{c}=2.474$ for a field applied along the $c$-axis. Concerning
anisotropy, we adopt the KSEA limit ($\kappa=0$) and thus the only relevant
parameter is $\lambda$ which may be estimated from the observed spin rotation
by an angle $\Delta\theta \equiv 2 \pi \zeta$ over a distance 
$d=a/{\sqrt{2}}$ along the $x$-axis. The incommensurability parameter
$\zeta$ is related to the period $L$ of Eq. (\ref{eq:3.9}) by
$\zeta = {\varepsilon}/L$, where $\varepsilon$ is the scale parameter
introduced in Eq. (\ref{eq:2.9}). One may actually choose the free 
parameter
$\varepsilon$ as $\varepsilon = D/J$ and thus $\lambda \equiv 1$ and
$ {\gamma}^{2}={\kappa} + {\lambda}^{2} + h^{2} = 1 + h^{2}$. At zero field, 
Eq. (\ref{eq:3.10}) is applied for $\lambda = 1 = \gamma$ to yield
${\delta}^{2} = 0.53189772$ and the period is calculated from Eq. 
(\ref{eq:3.9}) as $ L = 6.49945169$. Hence, $\varepsilon = {\zeta} L=0.1774$,
where we have also used the value $\zeta = 0.0273$ measured at zero
field \cite{5}. To summarize, our final choice of constants is
\begin{eqnarray}
\label{eq:3.16}
\kappa &=& 0 ,  \hspace*{2cm} \lambda \equiv 1 , 
              \\
              \nonumber
{\gamma}^{2}&=& 1 + h^{2} ,  \hspace*{1cm}   \varepsilon = D/J = 0.1774, 
\end{eqnarray}
and should be completed with the stipulation that the unit of field ($h=1$)
corresponds to $2s{\sqrt 2}{\varepsilon}J/g_{c}{\mu}_{B}=1.682$ T,
while the unit of frequency (energy) is $2s{\sqrt 2}{\varepsilon}J = 0.241$ 
meV. The magnetization per Cu atom is given by Eqs. (\ref{eq:3.14}) and
(\ref{eq:3.15}) in units of $sg_{c}{\mu}_{B} = 1.237 {\mu}_{B}$.
Distance is measured in units of $d/{\varepsilon} = 33.75 $ \AA.
\begin{figure}
\centerline{\hbox{\psfig{figure=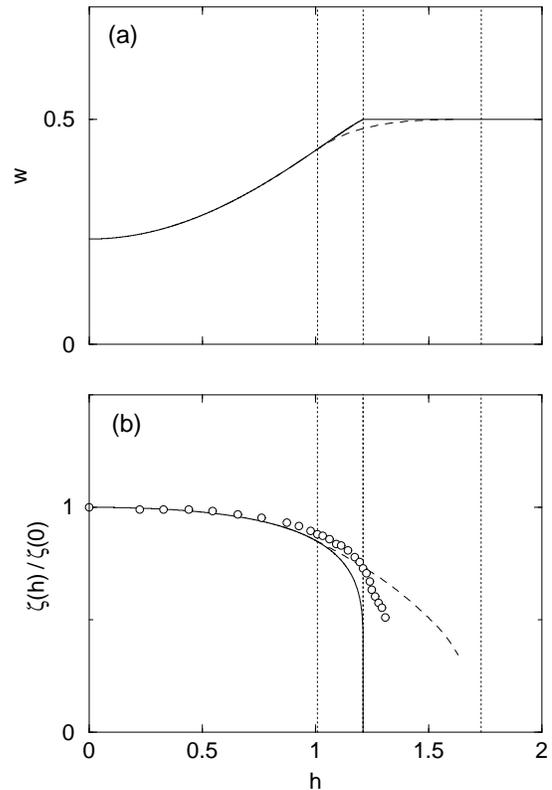,width=7.cm,bbllx=65bp,bblly=220bp,bburx=400bp,bbury=722bp}}}
\vspace*{0.5cm}
\caption { $T=0$ theoretical predictions for the
 field dependence of (a) the energy density $w$ and
 (b) the incommensurability parameter $\zeta$. Solid lines correspond
 to the flat spiral constructed in Sec. III and dashed lines to the
 nonflat spiral calculated in Sec. VI. The three vertical dotted lines
 indicate the location of the three critical fields $h_{1}=1.01$,
 $h_{c}=1.21$ and $h_{2}=1.73$, measured in units of $ 1.68$ T.
 Experimental data (open circles) measured at $T=2.4$ K were extracted 
 from Fig. 4 of Ref. 4.}
\end{figure} 

The constants (\ref{eq:3.16}) are inserted in Eq. (\ref{eq:3.12}) to yield
a critical field $h_{c}=1.21$ in rationalized units or $H_{c}=2.04$ T in
physical units. This theoretical prediction is consistent with
experiment and is 
thought to be a good indication that the KSEA
limit ($\kappa = 0$) may provide an accurate description of anisotropy
\cite{5}. Now, Eqs. (\ref{eq:3.10}) and (\ref{eq:3.9}) are applied with 
$\lambda = 1$ and ${\gamma}^{2}= 1 + h^{2}$ to yield the root 
$\delta = {\delta}(h)$ and the period $ L = L(h)$ at field $h$.
The field dependence of the energy density  computed from
Eq. (\ref{eq:3.11}) for $h<h_{c}$, and $w=1/2$ for $h>h_{c}$, is depicted
by a solid line in Fig. 3a. Similarly, the
field dependence of the incommensurability parameter
$\zeta = {\zeta}(h)$ is calculated from

\begin{equation}
\label{eq:3.17}
  \frac{{\zeta}(h)}{{\zeta}(0)} = \frac {L(0)}{L(h)} ,
\end{equation}
where ${\zeta}(0)$  and $L(0)$ are the zero-field parameters
already discussed, and is depicted by a solid line in Fig. 3b.
The results of Fig. 3 will be completed and further discussed in
Sec. VI. The same numerical data may be employed in Eqs. (\ref{eq:3.14}) and 
(\ref{eq:3.15}) to calculate the field dependence of the magnetization and 
the corresponding susceptibility. 

Finally, we return to the $U(1)$ transformation (\ref{eq:3.4}) which may be 
applied to the special solution (\ref{eq:3.6}) to yield a family of 
degenerate ground-state configurations:
 \begin{equation}
\label{eq:3.18}
 \Theta = {\theta}(x\ {\cos}{\psi}_{0} + y\ {\sin}{\psi}_{0}), \hspace*{1cm}
           \Phi = - {\psi}_{0},
\end{equation}
where ${\psi}_{0}$ is an arbitrary angle. The propagation vector of the
resulting spiral forms an angle ${\psi}_{0}$ with the $x$-axis, while
the normal to the spin plane forms an angle ${\pi}/2 - {\psi}_{0}$
with the same axis. For the special rotation ${\psi}_{0}={\pi}/4$, the
magnetic propagation vector and the normal to the spin plane are parallel
(screw-type spiral).
This symmetry operation is the basis for the bisection rule
discovered by Zheludev et al. \cite{3} when the external field is applied
in a direction \emph{perpendicular} to the $c$-axis, at an angle ${\chi}_{0}$
with respect to the $x$-axis. The normal to the spin plane rotates
almost freely to align with the external field, and thus 
${\chi}_{0} = {\pi}/2 - {\psi}_{0}$, in order to minimize (eliminate)
the positive term $({\bf n}\cdot{\bf h})^{2}$ in the potential (\ref{eq:2.15}).
The new term $({\bf h} \times {\bf d}_{z}) \cdot {\bf n}$ in the above 
potential does not affect the bisection rule but it does modify the profile
of the spiral. For example, when the field is applied along the $y$-axis
(${\chi}_{0} = {\pi}/2$, ${\psi}_{0}=0$, 
${\bf h} \times {\bf d}_{z} = {\lambda}^{\prime}h{\bf e}_{1}$) the staggered
magnetization is again confined in the $xz$-plane but the potential 
(\ref{eq:3.7}) becomes
\begin{equation}
\label{eq:3.19}
V = \frac{1}{2}\left[ \left({\theta}^{\prime} - \lambda \right)^{2}
       +{\gamma}^{2}{\cos}^{2}\theta \ \right] 
       + {\lambda}^{\prime}h\ {\sin}{\theta},
\end{equation}
where ${\gamma}^{2} = {\kappa} + {\lambda}^{2}$ is now field independent.
Nevertheless, the external field reappears in a different form and 
requires a new calculation of the spiral based on Eq. (\ref{eq:3.19}). Such
a calculation might actually explain the observed (weak) field dependence
of the magnitude of the magnetic propagation vector \cite{3} and provide
an estimate for the strength ${\lambda}^{\prime}$ (or $D^{\prime}$) of
the oscillating component of the DM anisotropy.

\section{Spin-flop phase}
\label{sec:spin_flop_phase}

We now begin to address questions of dynamics based on the complete
Lagrangian ${\cal L}={\cal L}_{0} - V$ of Eq. (\ref{eq:2.15}) applied for 
a field ${\bf h}=(0,0,h)$. If we also insert the spherical parameters 
(\ref{eq:3.5}) we find that

\begin{equation}
\label{eq:4.1}
{\cal L}_{0} =\frac{1}{2} \left(\dot{\Theta}^{2} 
          + {\sin}^{2}{\Theta}\ \dot{\Phi}^{2}\right)
          + h\ {\sin}^{2}{\Theta}\ \dot{\Phi} ,  
\end{equation}
and
\begin{eqnarray}
\label{eq:4.2}
      V =&& \frac{1}{2}
            \left[\left(\nabla {\Theta}\right)^{2} 
          + {\sin}^{2}{\Theta}\left(\nabla {\Phi}\right)^{2}
          +{\gamma}^{2}{\cos}^{2}{\Theta} + {\lambda}^{2}\right]    
      \nonumber
      \\
      && + {\lambda}\ \left[{\cos}{\Theta}\ {\sin}{\Theta}\ 
          \left({\sin{\Phi}}\ {\partial}_{1}{\Phi}
          +{\cos{\Phi}}\ {\partial}_{2}{\Phi}\right) \right.
      \nonumber
      \\
      && \left. - \cos{\Phi}\ {\partial}_{1}{\Theta}
          +\sin{\Phi}\ {\partial}_{2}\Theta \right] ,
\end{eqnarray}
where ${\nabla}=({\partial}_{1},{\partial}_{2})$ is the usual 2D gradient
operator, while the Laplacian will be denoted in the following
by ${\Delta} = {\partial}^{2}_{1} + {\partial}^{2}_{2}$.

We first study the high-field commensurate phase ($h > h_{c}$) where the
absolute minimum of the classical energy is thought to be the uniform
spin-flop state ${\bf n}= (1,0,0)$ or ${\Theta}={\pi}/2$ and ${\Phi}=0$.
Small fluctuations around this state are calculated by introducing
${\Theta}={\pi}/2 + f$ and ${\Phi}=g$ in Eqs. (\ref{eq:4.1}) and (\ref{eq:4.2})
and keeping terms that are at most quadratic in the small amplitudes
$f=f(x,y,{\tau})$ and $g=g(x,y,{\tau})$. Linear terms do not appear
because we are expanding around a stationary point of the energy functional,
whereas constants and total derivatives can be omitted because they do not
contribute to the equations of motion. Thus 
the corresponding linearized equations are found to be
\begin{equation}
\label{eq:4.3}
\ddot{f} - {\Delta}f + {\gamma}^{2}f = 2 {\lambda} {\partial}_{2} g ,
\hspace*{1cm} \ddot{g} - {\Delta}g = -  2 {\lambda} {\partial}_{2} f .
\end{equation}
Performing the usual Fourier transformation with frequency $\omega$
and wave vector $ {\bf q}=(q_{1},q_{2})$ one obtains a homogeneous system
whose solution requires that the corresponding determinant vanish. This
condition leads to two branches of eigenfrequencies

\begin{equation}
\label{eq:4.4}
{\omega}_{\pm}({\bf q}) =\left[ q^{2}_{1} + q^{2}_{2}
     +\frac {1}{2}\left( {\gamma}^{2} \pm 
     \sqrt{{\gamma}^{4} + 16 {\lambda}^{2}q^{2}_{2}}\right) \right]^{1/2} ,
\end{equation}
which will be referred to as the optical or acoustical mode, corresponding
to the plus or minus sign, respectively.

A notable feature of the calculated dispersions is their strong anisotropy.
In particular, the low-${\bf q}$ acoustical branch reads

\begin{equation}
\label{eq:4.5}
{\omega}_{-}({\bf q}) \approx \left[q^{2}_{1} + 
       \left(1 - 4{\lambda}^{2}/{\gamma}^{2}\right)q^{2}_{2}\ \right]^{1/2}
\end{equation}
and demonstrates that the spin-wave velocity depends on the direction
of propagation. It also makes it clear that an instability arises when
${\gamma}^{2} < 4{\lambda}^{2}$. In fact, the complete acoustical frequency
of Eq. (\ref{eq:4.4}) becomes purely imaginary over a nontrivial region
in ${\bf q}$-space when 
${\gamma}^{2}={\kappa} + {\lambda}^{2} + h^{2} < 4{\lambda}^{2}$.
Therefore, the uniform spin-flop state  is unstable for $h < h_{2}$ where

\begin{equation}
\label{eq:4.6}
  h_{2} = \sqrt{3{\lambda}^{2} - {\kappa}}
\end{equation}
is a new critical field. For our choice of parameters (\ref{eq:3.16})
$h_{2}=\sqrt{3}$ or $H_{2} = 1.682 \sqrt{3} = 2.91 $ T. The important
conclusion is that the spin-flop state is locally stable only for
$H > H_{2} > H_{c}$.

It is also interesting to examine the gap of the optical branch at ${\bf q}=0$
where 
${\omega}_{+}({\bf q}=0)= {\gamma} = ({\kappa} + {\lambda}^{2} + h^{2})^{1/2}$.
This result may be used to illustrate our earlier claim concerning the role
of the scale parameter $\varepsilon$. If we recall the definition of the 
rescaled parameters (\ref{eq:2.11}) and also include the factor 
$2s\sqrt{2}{\varepsilon}J$ to account for the physical unit of frequency, the
calculated gap is independent of $\varepsilon$ and is expressed entirely
in terms of constants that appear in the original discrete Hamiltonian of
Sec. II. Hence, in the KSEA limit, we find that
\begin{equation}
\label{eq:4.7}
  {\omega}_{+}({\bf q}=0) = \left[\left(2s\sqrt{2}D\right)^{2} 
  + \left(g_{c}{\mu}_{B}H\right)^{2}\right]^{1/2},
\end{equation}
in agreement with the magnon gap given in Ref. 5. Incidentally, this
special result is the only feature of the spectrum actually calculated
in the above reference for nonzero field.

The complete dispersions are illustrated in Fig. 4 for $H=3$ T, and for
spin-wave propagation along the $x$- or the $y$-axis.
\begin{figure}
\centerline{\hbox{\psfig{figure=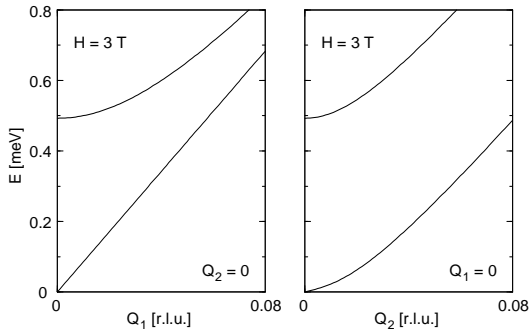,width=7.cm}}}
\vspace*{0.5cm}
\caption{ Theoretical magnon dispersions in the spin-flop phase,
 for spin-wave propagation along the $x$-axis ($Q_{2}=0$) and
 the $y$-axis ($Q_{1}=0$). }
\end{figure}
The anisotropy of the spectrum is made especially apparent by the fact that 
the dispersion of the acoustical mode is strictly linear in the $x$-direction,
but almost ferromagnetic-like in the $y$-direction because the chosen field
is only slightly greater than the critical field $H_{2} \approx 2.9$ T.
The numerical data for Fig. 4 were obtained from Eq. (\ref{eq:4.4}) applied
for our choice of units and constants given in Eq. (\ref{eq:3.16}). Thus
we set ${\lambda}=1$ and ${\gamma}^{2} = 1 + h^{2}$, with
$h= 3/1.682 = 1.784$, and also include an overall factor $0.241$ meV to account
for the physical unit of energy. Finally, ${\bf Q} = {\varepsilon}{\bf q}$
is the wave vector defined on the complete square lattice formed
by the Cu atoms within each layer, while relative units are defined from
$Q[r.l.u.]=Q/2{\pi} = ({\varepsilon}/2{\pi})q$. Therefore, Eq. (\ref{eq:4.4}) 
is applied with $q=(2{\pi}/{\varepsilon})\ Q[r.l.u.] = 35.418\ Q[r.l.u.].$

Unfortunately, there seem to exist no experimental data in the field region
$H \gtrsim 3$ T. In fact, the only published data \cite{5} were obtained
for $H = 2.5$ T $<H_{2}$ and spin-wave propagation along the $x$-axis.
For this special direction ($q_{2}=0$) the theoretical dispersions 
(\ref{eq:4.4}) do not ``see'' the instability. One may then deliberately 
apply them for $H=2.5$ T and compare the results to the actual data, 
as is done in Fig. 5 where a systematic disagreement is apparent in both 
dispersions.
In particular, the numerical fits to the data represented by dashed lines 
indicate a  significant 20\% reduction in the measured spin-wave velocity; 
as was already noted in Ref. 5.
\begin{figure}
\centerline{\hbox{\psfig{figure=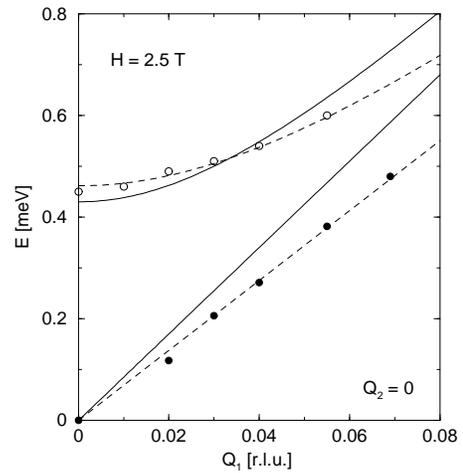,width=6cm}}}
\vspace*{0.5cm}
\caption{ Theoretical magnon dispersions (solid lines) in the spin-flop phase,
  deliberately applied for $H=2.5$ T $< H_{2}$. Circles denote experimental
 data \cite{5} and dashed lines are numerical fits to the same data.}
\end{figure}

Of course, our earlier discussion makes it clear that the dispersions 
(\ref{eq:4.4}) cannot be applied for $H = 2.5$ T because the corresponding
ground state is predicted to be unstable. At best, the fully polarized
spin-flop state ${\bf n}=(1,0,0)$ survives in the field region $H < H_{2}$ as
a metastable state thanks to some small tetragonal anisotropy that may be 
present in the discrete system \cite{3} but drops out of the leading continuum 
approximation. An appealing scenario suggested by our calculation is that
the system actually enters a different (intermediate) phase for $H < H_{2}$
which consists of some sort of mixed domains with no definite axis
of polarization.
Such a picture could explain the effective reduction 
of the spin-wave velocity, also taking into account the anisotropy of the
acoustical mode. 

As mentioned already, the continuum model does not contain anisotropies
that would necessarily polarize the staggered magnetization along
the $x$ (or the $y$) axis. Instead, there is a family of degenerate spin-flop
states ${\bf n} = (\cos{\Phi}_{0}, \sin{\Phi}_{0},0)$ with the same
energy for any constant angle ${\Phi}_{0}$. The corresponding small
fluctuations are now studied by introducing ${\Theta} = {\pi}/2+f$ and
${\Phi} = {\Phi}_{0}+g$ in Eqs. (\ref{eq:4.1}) and (\ref{eq:4.2}). A short
calculation similar to the one presented for $ {\Phi}_{0} = 0$ leads to
the magnon dispersions
\begin{equation}
\label{eq:4.8} 
\omega_{\pm}({\bf q})=
\left[ \, 
	{\bf q}^2 + \frac{1}{2} 
	\left(
		\gamma^{2} \pm 
		\sqrt{
			\gamma^{4} + 16 \lambda^{2}({\bf e}\cdot {\bf q})^{2}}
	\right) \, 
\right]^{1/2} ,
\end{equation}
where ${\bf e} = {\sin}{\Phi}_{0}\ {\bf e}_{1} + {\cos}{\Phi}_{0}\ {\bf e}_{2}$
is the unit vector obtained by rotating ${\bf e}_{2}$ with an angle 
$-{\Phi}_{0}$. The emerging picture is yet another manifestation
of the peculiar nature of the $U(1)$ symmetry (\ref{eq:3.4}), in some respects
similar to the bisection rule discussed in the concluding paragraph 
of Sec. III. In any case, the main conclusion of the present section
persists; namely, the acoustical mode develops maximum instability
along the direction ${\bf e}$ and leads to the same critical field given
earlier in Eq. (\ref{eq:4.6}).

The nature of the intermediate phase
will be discussed in Sec. VI.
The present section is concluded with a word of caution
concerning the validity of the continuum approximation at nonzero field,
which roughly requires that $g_{c}{\mu}_{B}H \ll J$. This strong inequality
becomes increasingly marginal for field strengths in the region
$H \gtrsim H_{2}$.
      
\section{Spiral phase}
\label{sec:spiral_phase}

The calculation of the low-energy magnon spectrum in the spiral phase
($h<h_{c}$) is significantly more complicated, but the general strategy
is identical to that followed in Sec. IV. Hence we introduce new fields
according to 

\begin{equation}
\label{eq:5.1}
       \Theta = \theta + f, \hspace*{1cm} \Phi = \frac{g}{\sin{\theta}},
\end{equation}
where $\theta = {\theta}(x)$ is the profile of the ground-state spiral
given by Eq. (\ref{eq:3.8}) while $f = f(x,y,\tau)$ and $g = g(x,y,\tau)$
account for small fluctuations. The special rescaling chosen in the
second equation is equivalent to working in a \emph{rotating frame} \cite{18}
whose third axis is everywhere parallel to the direction of the background
staggered magnetization ${\bf n}=(\sin{\theta},0,\cos{\theta})$.

The new fields (\ref{eq:5.1}) are introduced in the complete Lagrangian given 
by Eqs. (\ref{eq:4.1}) and (\ref{eq:4.2}) which is then expanded to second 
order in $f$ and $g$. The required algebra is lengthy but
the final result for the linearized equations is
sufficiently simple:

\begin{eqnarray}
\label{eq:5.2}
  \ddot{f} - {\Delta}f + U_{1}f &=& 2h \cos{\theta}\ \dot{g}
                    + 2 {\lambda}\sin{\theta}\ {\partial}_{2} g,
              \nonumber
              \\
\ddot{g} - {\Delta}g + U_{2}\ g &=& - 2h \cos{\theta}\ \dot{f}
                             - 2 {\lambda}\sin{\theta}\ {\partial}_{2} f,
\end{eqnarray}
where

\begin{eqnarray}
\label{eq:5.3}
  U_{1} &=& -\ {\gamma}^{2} \cos (2{\theta}),
                          \nonumber
                          \\
  U_{2} &=& 2 {\lambda} \sqrt{{\delta}^{2} + {\gamma}^{2} {\cos}^{2}{\theta}}
            -2 {\gamma}^{2} {\cos}^{2}{\theta} - {\delta}^{2},
\end{eqnarray}
are effective potentials that can be calculated for any desired set of 
parameters, as explained in Sec. III. The general idea that the calculation 
of the spectrum in a spiral antiferromagnet can be reduced to 
a Schr\"odinger-like  problem in a periodic potential is not new
\cite{19}, but the specific structure of Eqs. (\ref{eq:5.2}) requires special
attention.

We found it instructive to consider first the special case of spin-wave 
propagation along the $x$-axis (${\partial}_{2} f = 0 = {\partial}_{2} g$)
at zero external field ($h=0$). This is actually the only case for which
the low-energy spectrum was previously calculated starting from the discrete
Hamiltonian \cite{5}. If we further perform the temporal Fourier transformation
with frequency $\omega$, Eqs. (\ref{eq:5.2}) reduce to
\begin{equation}
\label{eq:5.4}
     - f^{\prime \prime} + U_{1}f = {\omega}^{2}f, \hspace*{1cm}
     - g^{\prime \prime} + U_{2}g = {\omega}^{2}g,
\end{equation}
where the prime denotes differentiation with respect to $x$. Therefore,
in this special case, the eigenvalue problem is reduced to two decoupled
1D Schr\"odinger  equations of the standard type with potentials
$U_{1}$ and $U_{2}$ calculated at zero field. Also note that both potentials
are periodic functions of $2{\theta}$ and thus their period is actually
$L/2$ where $L$ is the period of the background spiral.

The eigenvalue problems (\ref{eq:5.4}) are solved in Appendix A.
The numerical procedure yields eigenfrequencies
$\omega = {\omega}(q_{1})$ as functions of Bloch momentum $q_{1}$. 
The latter can be restricted to the zone $[-2{\pi}/ L , 2{\pi}/ L]$,
because the period of the potentials is $L/2$, or to the zone
$[-{\zeta},{\zeta}]$ in relative units defined as in Sec. IV. Several
low-lying eigenvalues are illustrated in Fig. 6a using a reduced-zone scheme.
Solid and dashed lines correspond to the first and second eigenvalue problem
in Eq. (\ref{eq:5.4}) and are superimposed in the same graph for convenience.
We also find it convenient to refer to the two types of modes as acoustical
and optical. In either case, there is only one discernable gap that occurs
between the first and the second band at the zone boundary. The calculated
boundary gaps are $0.123$ meV and $0.049$ meV, respectively, while the absolute
gap of the optical mode at the zone center is $0.170$ meV. All of the above
theoretical predictions agree with those obtained in Ref. 5 by
a different method. They also agree with experiment, except for the small
($0.049$ meV) gap that has not yet been resolved at zero field.

The same results are depicted in Fig. 6b using an extended-zone scheme.
In fact, this figure displays two replicas of the acoustical mode centered
at $\pm {\zeta}$. The need for two replicas follows from the structure
of dynamic correlation functions in the laboratory frame, rather than in the
rotating frame actually used in the calculation of the magnon spectrum 
\cite{5}. Our results in Fig. 6b are obviously consistent with both
the experimental and the theoretical results obtained in the above reference
at zero field.
\begin{figure}
\centerline{\hbox{\psfig{figure=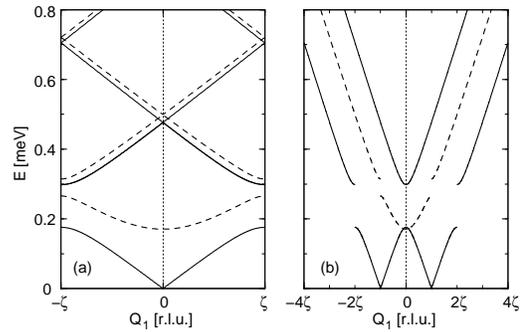,width=7.cm}}}
\vspace*{0.5cm}
\caption{Magnon spectrum for spin-wave propagation along the $x$-axis
 at zero field. Solid and dashed lines distinguish between acoustical
 and optical modes.
 (a) The spectrum in a reduced-zone scheme, and 
 (b) the same spectrum in an extended-zone scheme including two
 replicas of the acoustical mode centered at $\pm {\zeta}$.}
\end{figure}

We are now in a position to extend the calculation to the general case
of nonzero field and arbitrary direction of spin-wave propagation. 
The external field enters Eqs. (\ref{eq:5.2}) in two distinct ways. First,
it affects the structure of the potentials $U_{1}$ and $U_{2}$ because
the background spiral is further distorted. Second, the field induces
first-order time derivatives which originate in the ``nonrelativistic''
term of Eq. (\ref{eq:4.1}) and couple the two linear equations (\ref{eq:5.2}).
Additional coupling between the two equations appears in the case
of arbitrary direction of propagation because ${\partial}_{2}f$ and
${\partial}_{2}g$ no longer vanish. Altogether we are faced with
a nonstandard eigenvalue problem that is also solved in Appendix A.

Here we present explicit results for four typical
values of the rationalized field $h=0$, $0.3$, $0.6$ and $0.9$ which will be 
quoted from now on by their rounded physical values 
$H=0$, $0.5$, $1$ and $1.5$ T.
In Fig. 7 we illustrate the calculated spectrum for spin-wave propagation 
along the $x$-axis ($q_{2}=0$) using a highly reduced zone scheme.
An important check of consistency is provided by the fact that the $H=0$
results of Fig. 7 agree with those presented earlier in Fig. 6a, except
that the zone is now reduced down to [$-{\zeta}/2,{\zeta}/2$]
for reasons explained in Appendix A. Furthermore,
we no longer employ solid and dashed lines to distinguish between
acoustical and optical modes. Such a distinction is not a priori possible
in the current algorithm because of the coupling (hybridization) of the
two types of modes at nonzero field.
\begin{figure}
\centerline{\hbox{\psfig{figure=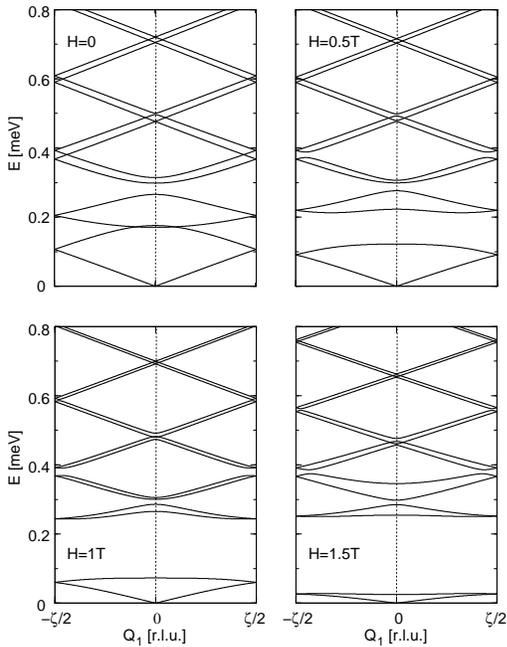,width=7.cm}}}
\vspace*{0.5cm}
\caption{ Magnon spectra for spin-wave propagation along the $x$-axis, at four 
 values of the applied field $H$, using a highly reduced
 zone scheme.}
\end{figure} 

One should keep in mind that the extent of the zone [$-{\zeta}/2,{\zeta}/2$]
slides with the applied field, a feature that is not apparent in Fig. 7
because the scale of the abscissa is adjusted accordingly. The 
incommensurability parameter $\zeta=0.0273$ measured at $H=0$ is used
as input in our calculation. The calculated values for $H=0.5$, $1$ and
$1.5$ T are $\zeta=0.0271$, $0.0264$ and $0.0245$. 

At first sight, it would seem difficult to extract useful information
from the highly convoluted spectra shown in Fig. 7.
Nevertheless, the most
vital information concerning the low-energy dynamics is easily abstracted
from Fig. 7 because the low-lying bands are clearly segregated. In particular,
it is still possible to distinguish between the acoustical and the optical
mode, at least in an operational sense. Thus we unfold the first six branches
back to the zone [$-{\zeta},{\zeta}$] and then proceed to the extended-zone
scheme of Fig. 6b including two replicas of the acoustical mode centered
at $\pm {\zeta}$. The resulting low-energy spectra are shown in Fig. 8.
\begin{figure}
\centerline{\hbox{\psfig{figure=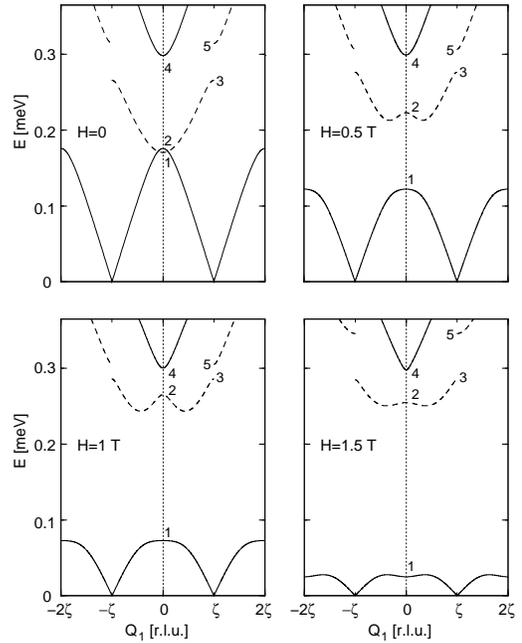,width=7.cm}}}
\vspace*{0.5cm}
\caption{The low-energy spectra of Fig. 7 using an extended-zone scheme
 as in Fig. 6b. The energy values at the five characteristic spectral
 points denoted by 1, 2, 3, 4 and 5 are given in Table I.}
\end{figure}

The $H=0$ entry of Fig. 8 is but a magnified version of the lower-central
portion of Fig. 6b, as expected. This version reveals a certain ``anomaly''
that is not conspicuous in Fig. 6b, namely, a relative crossing between
the two modes in a narrow region around the zone center. The calculated
maximum splitting of $0.005$ meV is within the error margin of the continuum
approximation and, in any case, beyond experimental detection. But
the resolution of this theoretical curiosity is interesting: when the direction
of spin-wave propagation departs slightly from the $x$-axis ($q_{2} \neq 0$)
and/or a finite field is turned on, the crossing points become avoided
crossings. Therefore, strictly speaking, the solid and dashed lines must
be interchanged in the narrow region between the two crossing points.
This explains the apparent slight inconsistency in the labeling of the
five characteristic points of the spectrum denoted by 1, 2, 3, 4 and 5 in
Fig. 8. The calculated magnon energies at those points are summarized
in Table I.
\begin{table}
\caption{Energy in units of meV at the five characteristic points
 of the spectrum denoted by 1, 2, 3, 4 and 5 in Fig. 8.}
\label{table:I}
\begin{tabular}{cccccc}
 $H$ [T] &  $E_{1}$ &  $E_{2}$ & $E_{3}$ & $E_{4}$ & $E_{5}$ \\
\hline
 $0.0$   &  $0.170$ &  $0.176$ & $0.266$ & $0.298$ & $0.314$ \\
 $0.5$   &  $0.122$ &  $0.223$ & $0.276$ & $0.299$ & $0.307$ \\
 $1.0$   &  $0.073$ &  $0.265$ & $0.286$ & $0.300$ & $0.305$ \\
 $1.5$   &  $0.025$ &  $0.255$ & $0.285$ & $0.298$ & $0.346$ \\
\end{tabular}                                                                  \end{table}

We now concentrate on the optical mode. The gap $E_{2}=0.176$ meV calculated
at zero field agrees with the measured $0.18$($1$) meV. Our calculation
further shows that the above gap evolves quickly with increasing field
to reach the asymptotic value $0.26$ meV around which it oscillates
mildly. The complete optical mode evolves into a snake-like dispersion
with energy values in the range $0.25$ meV $< E < 0.29$ meV. These predictions
are generally consistent with experiment \cite{5}.
However, some of the finer details deserve closer attention. The calculated
energy at point 5 in the spectrum remains practically constant at
$E_{5} \approx 0.31$ meV for $H\lesssim 1$ T, while a steep crossover takes 
place for higher field values which leads to $E_{5} \approx 0.35$ meV
for $H=1.5$ T. These predictions are also in
agreement with experiment \cite{5}. But the calculated splittings
of the optical dispersion $E_{5} - E_{3} = 0.02$ meV and $0.06$ meV,
for $H=1$ and $1.5$ T, disagree with the measured 0.05 meV and 0.11 meV.
It appears that the observed splittings are better described by 
$E_{5} - E_{2} = 0.04$ meV and 0.09 meV. In fact, the above identification
may not be completely arbitrary. For instance, the lowest branch in the optical
dispersion measured for $H=1.5$ T shows a clear local maximum of 0.28 meV at
the zone center, which agrees with the calculated maximum $E_{3}=0.285$ meV
at the zone boundaries $\pm {\zeta}$ rather than the gap $E_{2}=0.255$ meV
at the zone center. It seems that the lowest branch in the observed
optical dispersion for $H=1.5$ T is composed of two replicas of the calculated
dispersion centered at $\pm {\zeta}$. On the other hand, experimental data
\cite{5} at higher energies not shown in Fig. 8 indicate the appearance
of two replicas centered at $\pm 2{\zeta}$. Unfortunately, we cannot resolve
this issue of proper replication of the basic modes because our current
formalism does not directly address the relevant dynamic correlation functions.

Next we discuss the acoustical mode. Our calculation shows that the energy
at point 4 in the spectra of Fig. 8 remains remarkably stable at 
$E_{4} \approx 0.30$ meV for all field values considered. This feature is
also in agreement with experiment which indicates only a mild decline
from the above value with increasing field. Nevertheless, a clear
disagreement occurs in the lowest branch of the acoustical mode. Although
explicit data points are not given for this branch by Zheludev et al. 
\cite{5}, the solid lines in their Figs. 6 and 7, and the corresponding
wording in their text, suggest that the lowest branch in the measured 
spectrum is also largely insensitive to the applied field. In contrast,
our calculation predicts a robust reduction of the energy gap $E_{1}$
with increasing field (see Fig. 8 and Table I). The calculated
spin-wave velocity is also reduced, albeit at a slower rate.

The preceding apparent disagreement with experiment is especially important
because it is directly related to the issue of local stability
of the spiral phase. Indeed, a careful numerical investigation reveals
that the gap $E_{1}$ vanishes at the critical field $h_{1} \approx 1.01$, or
$H_{1} \approx 1.70$ T, while an unstable mode develops for $H > H_{1}$.
This mode is first detected by the appearance of a \emph{real} eigenvalue
in the matrix $M$ of Eq. (A5),
when $H$ crosses $H_{1}$, which corresponds to
purely imaginary frequency. As the field increases beyond $H_{1}$ the
instability occurs over a nontrivial region in ${\bf q}$-space. Therefore,
the flat spin spiral constructed in Sec. III is predicted to be
locally stable only for $H < H_{1} < H_{c}$.

It is interesting that the experimental work \cite{4,5} already provided
evidence for the existence of a critical field $H_{1} = 1.7$ T that
coincides with our theoretical prediction. However, one should also 
contemplate the possibility that such a coincidence may be 
fortuitous, in view of the apparent contradiction between experimental
and theoretical predictions for the gap $E_{1}$. In any case, our current
result together with the discussion of Sec. IV clearly suggest the existence
of an intermediate phase in the field region $1.7$ T $< H < 2.9$ T.
The nature of the
intermediate phase is discussed in Sec. VI.

In the remainder of this section we take a different view
of the low-energy magnon spectrum by considering spin-wave
propagation along the normal to the plane of the flat spiral.
Our algorithm is adapted to this case simply by setting the Bloch wave number
$q_{1}=0$ and calculating frequencies as functions of the wave number $q_{2}$
in the $y$-direction. It is interesting that no theoretical or experimental
results exist in this case even at zero field. Our results are illustrated
in Fig. 9 for the same set of field values employed in the preceding 
discussion.
\begin{figure}
\centerline{\hbox{\psfig{figure=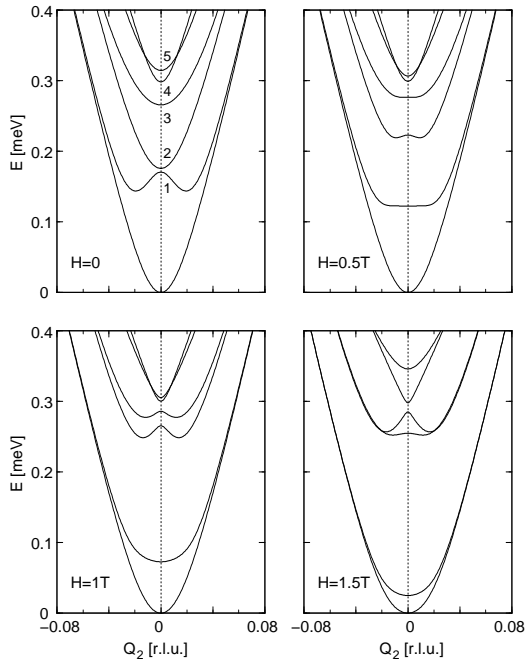,width=7.cm}}}
\vspace*{0.5cm}
\caption{ Magnon spectra for the same field values as in Figs. 7 and 8        
 but spin-wave propagation along the $y$-axis.}
\end{figure}

The most stable feature of Fig. 9 is its lowest branch which exhibits
\emph{quadratic} dependence on $q_{2}$ near the origin. Clearly this
branch is the extension of the acoustical dispersion in the 
$y$-direction originating at its points where $E = 0$. Therefore, the complete
acoustical mode is Goldstone-like in the $x$-direction but ferromagnetic-like
in the $y$-direction. Such a characteristic anisotropy is in some
respects similar to the situation encountered in the spin-flop phase
discussed in Sec. IV.

Higher branches labeled as 1, 2, 3, 4 and 5 in Fig. 9 also possess
a simple interpretation, for they are the extensions in the $y$-direction
of the special spectral points numbered accordingly in our earlier Fig. 8.
In contrast to the fundamental ferromagnetic-like branch, higher branches
evolve vigorously with the applied field. In particular, branch 1 in Fig. 9
is quickly depressed with increasing field to become degenerate
with the fundamental branch at the critical field $H_{1} = 1.70$ T
not included in the figure. For $H > H_{1}$ this mode becomes unstable
over a nontrivial region of wave numbers around the origin. Of course,
this is the instability described earlier in the text viewed from
a different perspective.

We have thus provided a fairly complete theoretical picture of the
low-energy magnon spectrum, including predictions for which there exist
no experimental data at present. It is interesting to see whether or not
future experiments could resolve the apparent discrepancy in the field
dependence of the magnon gap $E_{1}$, and thus illuminate the important
issue of local stability of the spiral phase, as well as confirm the
predicted characteristic anisotropy in the low-energy spectrum.

\section{Intermediate phase}
\label{sec:intermediate_phase}

We now focus on the predicted intermediate phase and examine its
nature through a direct numerical minimization of the complete
energy functional $W$ of Eqs. (\ref{eq:3.1}) and (\ref{eq:3.2}).
The method of calculation is a relaxation algorithm
formulated on the basis of a discretized form of the energy
functional defined on a square grid. After long experimentation
with 2D simulations, it progressively became apparent that the
optimal configuration for $h>h_{1}$ is actually a 1D nonflat spiral
characterized by a staggered magnetization whose three components
are all different than zero.

Therefore, an accurate calculation of the nonflat spiral was 
eventually obtained by a relaxation algorithm applied directly
to a 1D restriction of the energy functional whose variation
leads to the coupled stationary equations:
\begin{eqnarray}
\label{eq:6.1}
  & \Theta^{\prime\prime}+(\gamma^2-\Phi^{\prime 2})
       \cos\Theta\, \sin\Theta
    = -2\,\lambda\,\sin^2\Theta\,\sin\Phi\;\Phi^{\prime} &
    \nonumber
    \\
  & (\sin^2\Theta\, \Phi^{\prime})^{\prime}
    = 2\,\lambda\,\sin^2\Theta\,\sin\Phi\;\Theta^{\prime} &
\end{eqnarray}
These are ordinary differential equations because both angular
variables $\Theta$ and $\Phi$ are assumed to be functions of the
single coordinate $x$, while the prime again denotes differentiation
with respect to $x$. Nevertheless, it does not seem possible
to obtain analytical solutions of Eqs. (\ref{eq:6.1}), except for
the case of the flat spiral ($\Phi = 0$) discussed in Sec. III.
A significant obstacle is the fact that the period 
of the nonflat spiral is not known a priori. Hence our numerical
solution was carried out on a periodic 1D grid with specified
length $L$, until a relaxed configuration was obtained with 
energy density $w=w(L)$. We then varied $L$ to achieve the
least possible energy for each field $h$, and the corresponding
optimal period $L=L(h)$.

An important check of consistency is that the above algorithm
reproduces the results for the flat spiral obtained more directly
in Sec. III, but only when $h<h_{1}=1.01$. Instead, a nonflat spiral
emerges as the optimal solution for $h>h_{1}$. The calculated 
configuration is illustrated in Fig. 10 for a field value $h=1.21$
deliberately chosen to be equal to the critical field $h_{c}$ of the
conventional CI transition.
\begin{figure}
\centerline{\hbox{\psfig{figure=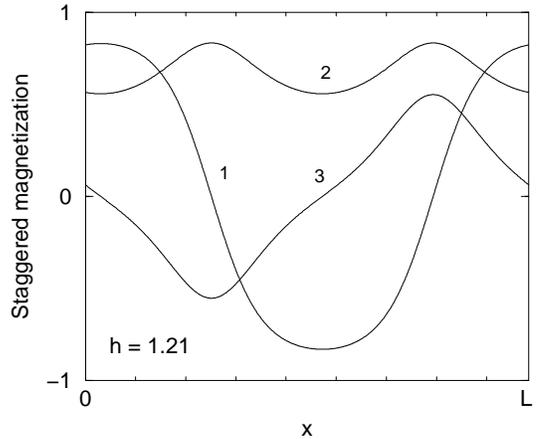,width=7.cm}}}
\vspace*{0.5cm}
\caption{ Profile of the nonflat spiral for $h=1.21$. The three
 curves correspond to the three components of the staggered magnetization
 $n_{1}$, $n_{2}$ and $n_{3}$. The calculated period is $L=8.84$.}
\end {figure} 
The energy of the nonflat spiral 
depicted by a dashed line in Fig. 3a is smaller than the energy
of both the flat spiral and the uniform spin-flop state throughout
the intermediate region $h_{1}<h<h_{2}$. One should also stress
that the nonflat spiral is here predicted to occur for a field 
applied strictly along the c-axis, and is not due to sample
misalignment \cite{4} or the presence of a transverse magnetic
field \cite{20}.

In a sense, the predicted intermediate phase smooths out 
the original sharp CI transition. This smoothing is also 
apparent in the calculated field dependence of the period $L=L(h)$
which is inserted in Eq. (3.17) to yield the results for the
incommensurability parameter shown by a dashed line in Fig. 3b.
The same figure displays experimental data taken from Ref. 4
where they were analyzed in terms of the conventional CI transition
based solely on a flat spiral. It should be noted that both the
measured zero-field incommensurability parameter $\zeta(0)=0.0273$
and an experimental critical field $H_{c}=2.15$ T  were used as
adjustable parameters in the theoretical analysis of Refs. 4 and 5
to obtain a reasonable overall fit. Yet the experimental data
indicate some smoothing of the CI transition near the critical field.
This fact is made apparent in our Fig. 3b where theoretical
results for both the flat spiral (solid line) and the nonflat
spiral (dashed line) are calculated using as input only the zero-field
parameters given earlier in Eq. (3.16).

Nevertheless, the results of Fig. 3b cannot be interpreted as
unambiguous evidence for the existence of an intermediate phase,
especially because the experimental data were taken at the 
relatively high temperature $T=2.4$ K. It is feasible that the
$T=0$ theoretical predictions could be further focused by invoking
deviation from KSEA anisotropy that is allowed by symmetry;
i.e., by repeating the calculation for nonzero values of the free
parameter $\kappa$. One should also keep in mind that a completely
accurate description of the CI transition may not be attainable
within the classical approximation.

The nonflat spiral exists as a stationary point of the energy
functional throughout the intermediate phase and degenerates
into a uniform spin-flop state polarized along the $y$-axis
near the upper critical field $h_{2}=1.73$. Actually, our
calculation was not pushed all the way to the critical field
$h_{2}$ because of numerical difficulties that occur as the
period grows to infinity. The theoretical
analysis should be completed with a detailed study of the stability
and dynamics of the nonflat spiral
within the full 2D context, in a manner analogous 
to our treatment of the flat spiral in Sec. V. The required
computational effort is too great to be included in the present
paper, especially because the profile of the nonflat spiral
is obtained numerically through the relaxation algorithm.
A future analysis could, in principle, reveal the existence
of yet another critical field within the intermediate region,
beyond which the nonflat spiral may cease to be locally stable.
It is thus important to also examine the nature of instability
at the upper critical field $h_{2}$, as discussed further in
Appendix B.

The configuration of Fig. 10 may be viewed as a conical spiral
that nutates around the $y$-axis. It is interesting that a
simple conical spiral without nutation had been discussed
theoretically in connection with the cholesteric-nematic 
transition in liquid crystals \cite{9,10} but has not yet
been observed experimentally because its realization requires
an anomalously small bend modulus \cite{11}. In contrast,
the parameters of Ba$_{2}$CuGe$_{2}$O$_{7}$ favor the
occurrence of the currently predicted intermediate phase.

\section{Conclusion}
\label{sec:conclusion} 

We have presented a field theoretical description of the low-energy
dynamics in the spiral antiferromagnet Ba$_{2}$CuGe$_{2}$O$_{7}$.
We have thus been able to calculate the low-energy magnon spectrum
for any strength of the applied field and any direction of spin-wave
propagation. In this respect, the present work significantly extends
the results of Ref. 5 where the spectrum was calculated only at zero
field and for propagation along the direction of the spiral.
Therefore, our theoretical results are relevant for the
analysis of experimental data obtained for nonzero field, which
were previously analyzed mostly in terms of empirical formulas.

An interesting byproduct of this detailed spin-wave analysis is the
identification of the two new critical fields $H_{1}$ and $H_{2}$,
and a corresponding prediction of an intermediate phase that does
not seem to be inconsistent with available experimental data.
The apparent discrepancy in the field dependence of the magnon gap
$E_{1}$ pointed out in Sec. V needs to be clarified, 
but could be due to poor experimental resolution at this rather
low energy scale ($0.1$ meV or less).
The field dependence of the incommensurability parameter 
discussed in Sec. VI could be
rectified by invoking a slight deviation from the KSEA
limit that is allowed by symmetry. Susceptibility
data \cite{4} taken at $T=2$ K display a rounded maximum 
which could be explained as a finite-temperature effect but
does not a priori exclude an intermediate phase. Furthermore, the set
of data for the magnon dispersion discussed in connection with Fig. 5 is
too limited to provide a clear picture. Therefore, a clear identification
or disproof of the intermediate phase may require additional
experimental work guided by the theoretical predictions 
of the present paper.

On the other hand, it is desirable to carry out a complete theoretical
analysis of the stability and dynamics of the intermediate phase along
the lines outlined in Sec. VI.
A related project is to extend our approach to the case of a field
applied in a direction perpendicular to the $c$-axis \cite{3}. The 
field-dependent modifications of the spiral can be computed on the basis
of Eq. (\ref{eq:3.19}), and a corresponding calculation of the low-energy
magnon spectrum can be carried out by a straightforward extension of
the methods developed in Sec. V.

Finally, we must comment on the two basic approximations made in the present
work. The adopted classical approach is equivalent to the usual semiclassical
approximation obtained by the $1/s$ expansion restricted to leading order.
The omitted quantum (anharmonic) corrections are not negligible in this
2D problem but are offset in part by the fact that the input parameters are 
consistently estimated within the classical approximation \cite{1,2,3,4,5}.
One should also question the validity of the continuum approximation
whose relative accuracy can be roughly estimated from 
 ${\varepsilon}^{2} \approx 0.03$ at zero field, but may deteriorate
in the presence of a strong external magnetic field.
Incidentally, the corresponding parameter
${\varepsilon}$ in a typical weak ferromagnet such as an orthoferrite
(YFeO$_{3}$) or a high-T$_{c}$ superconductor (La$_{2}$CuO$_{4}$) is at least
one order of magnitude smaller. In any case, the physical picture derived is 
sufficiently complete to provide a basis for a meaningful discussion of further
refinements.

\acknowledgements

We thank A. Bogdanov for bringing Refs. 20 and 21 to our
attention, S. Trachanas
for valuable suggestions concerning the eigenvalue problems studied
in the present paper, and M. Marder for a careful reading of the
manuscript. The work was supported in part by a Marie Curie 
Fellowship (HPMT-GH-00-00177-03), 
by a TMR program (ERBFMRXCT-960085), and by VEGA 1/7473/20.

\appendix
\section{Eigenvalue problems}
\label{sec:eigenvalue_problems}

The eigenvalue problems (\ref{eq:5.4}) were solved numerically, as explained
here for the first equation. Taking into account that the period of the
potential is $L/2$, the Bloch representation of the wave function reads

\begin{equation}
\label{eq:A1}
     f(x) =  e^{iq_{1}x} \sum_{n=-\infty}^{\infty}
             f_{n}\ exp(i4n{\pi}x/L)
\end{equation}
and the wave equation becomes

\begin{equation}
\label{eq:A2}
         (q_{1} +  4n{\pi}/ L)^{2}f_{n} +  
         \sum_{m=-\infty}^{\infty} U_{1,n-m}f_{m} = {\omega}^{2}f_{n},
\end{equation}
where the Fourier coefficients of the potential are given by
 
\begin{eqnarray}
\label{eq:A3}
   U_{1,n} &=& \frac {2}{L} \int_{-L/4}^{L/4} 
               exp(-i4n{\pi}x/ L)\ U_{1}[{\theta}(x)]\ dx
               \nonumber
               \\
           &=&\frac {4}{ L} \int_{0}^{{\pi}/2} 
               \cos \left[ {\frac {4n{\pi}}{L}} x({\theta}) \right]
\frac {U_{1}({\theta})\ d{\theta}}{\sqrt{ {\delta}^{2}+{\gamma}^{2}{\cos}^{2}
{\theta}  }} .
\end{eqnarray}
Here we use the fact that $U_{1}$ is an even function of $\theta$ or $x$,
and $x = x(\theta)$ is given by the integral (\ref{eq:3.8}). Thus the last step
of Eq. (\ref{eq:A3}) is in effect a double integral that is computed 
by an adaptive Newton-Cotes algorithm. The eigenvalue equation (\ref{eq:A2})
is then solved by diagonalizing the finite matrix that results from
a restriction of the indices $m$ and $n$ to the interval $[-N,N]$ where $N$
can be as low as 20. To be sure, only the first few Fourier coefficients
of the potential $U_{1}$ and $U_{2}$ are important, as demonstrated 
in Table II using as input the zero-field parameters quoted in Sec. III.
The numerical procedure just described yields eigenfrequencies
$\omega = {\omega}(q_{1})$ as functions of Bloch momentum $q_{1}$ that
can be restricted to the zone $[-2{\pi}/ L , 2{\pi}/ L]$,
or to $[-{\zeta},{\zeta}]$ in relative units defined as in Sec. IV.

\begin{table}
\caption{Fourier coefficients of the potentials $U_{1}$ and $U_{2}$
 at zero field. The table should be completed with the symmetry
 relations $U_{1,-n}=U_{1,n}$ and $U_{2,-n}=U_{2,n}$.}
\label{table:II}
\begin{tabular}{ccc}
    n   & $U_{1,n}$     & $U_{2,n}$         \\
\hline
   $0$  &\hspace*{0.33cm}$0.13034455$  &\hspace*{0.15cm} $0.53189772$      \\
   $1$  & $-0.49358342$ & $-0.24049378$     \\
   $2$  & $-0.06461030$ & $-0.04790238$     \\
   $3$  & $-0.00637043$ & $-0.00527215$     \\
   $4$  & $-0.00055833$ & $-0.00048614$     \\
   $5$  & $-0.00004588$ & $-0.00004113$     \\
   $6$  & $-0.00000362$ & $-0.00000331$     \\
   $7$  & $-0.00000028$ & $-0.00000026$     \\
   $8$  & $-0.00000002$ & $-0.00000002$    \\
\end{tabular}
\end{table}

We now return to the general case of nonzero field and arbitrary
direction of spin-wave propagation. We first rewrite Eqs. (\ref{eq:5.2})
in a form that contains only first-order time derivatives.
Hence we treat $u=\dot f$ and $v=\dot g$ as independent fields and introduce 
the four-component spinor $\cal X$ defined from ${\cal X}^{T}=(u,v,f,g)$.
Then Eqs. (\ref{eq:5.2}) read

\begin{equation}
\label{eq:A4}
 \dot{\cal X} = M {\cal X},
\end{equation}
where $M$ is the differential operator

\begin{eqnarray}
\label{eq:A5}
M &=& 
\left[\begin{array}{cccc}
 \hspace*{0.2cm}0    & 0        &\hspace*{0.1cm}   I  &   0      \\
 \hspace*{0.2cm}0    & 0        &\hspace*{0.1cm}   0  &   I      \\
 -D_{1}&\hspace*{0.2cm}D_{3}&\hspace*{0.1cm}   0  &   D_{4}\\
 -D_{3}&  -D_{2}            & -D_{4}&   0 \end{array}\right] . 
\end{eqnarray} 
Here $D_{1}=-{\Delta} + U_{1}$, $D_{2}=-{\Delta} + U_{2}$,
$D_{3}=2{\lambda}\sin{\theta}\ {\partial}_{2}$, 
$D_{4}=2h \cos {\theta}$, and $I$ is the unit operator. The chief advantage
of $M$ is that it does not contain time derivatives. A superficial
disadvantage is that $M$ is not a hermitian operator. In fact, Eq. 
(\ref{eq:A4}) suggests that the eigenvalues of $M$ are purely imaginary and
come in pairs $\pm i{\omega}$ where $\omega$ is the desired physical
frequency. A real eigenvalue in $M$ would correspond to purely
imaginary physical frequency and thus indicate instability of the
ground-state spiral. All of these features are explicitly realized
in the following numerical calculation.

Our task is then to construct a matrix representation of the differential
operator $M$. Attention should be paid to the fact that the Bloch
theorem must now be applied with the full period $L$ of the
spiral because of those terms in Eq. (\ref{eq:A5}) that are proportional
to $\cos{\theta}$ and $\sin{\theta}$. Hence the operator
$D_{1}=-{\Delta} + U_{1}$ is replaced by a matrix ($D_{1,nm}$)
with elements

\begin{equation}
\label{eq:A6}
   D_{1,nm} = \left[\left(q_{1} + 2n{\pi}/ L \right)^{2} 
              + q^{2}_{2}\right] {\delta}_{nm} + U_{1,n-m},
\end{equation}
where $q_{1}$ is now restricted to the zone $[-{\pi}/L,{\pi}/L]$, or
$[-{\zeta}/2,{\zeta}/2]$ in relative units, while $q_{2}$ is unrestricted
because the spiral depends only on $x$. Accordingly, the Fourier coefficients
of the potential are given by

\begin{equation}
\label{eq:A7}
  U_{1,n} = \frac {2}{ L} \int^{\pi}_{0}
            \cos \left[{\frac {2n{\pi}}{ L}}x({\theta}) \right]
\frac {U_{1}(\theta)\ d{\theta}}
{\sqrt{{\delta}^{2}+{\gamma}^{2}{\cos}^{2}{\theta}}},
\end{equation}
which differs from Eq. (\ref{eq:A3}) only in that the full period $L$,
instead of $L/2$, is employed. As a result, odd 
coefficients in Eq. (\ref{eq:A7}) vanish, while the collection of even
coefficients coincides with that obtained from Eq. (\ref{eq:A3}).
The operator $D_{2}$ is treated in exactly the same way replacing $U_{1}$ with
$U_{2}$. On the other hand, the operator 
$D_{3}=2{\lambda}\sin{\theta}\ {\partial}_{2}$ in Eq. (\ref{eq:A5}) is
replaced by $2{\lambda}q_{2}S$ where $S$ is an antisymmetric matrix
whose n-th codiagonal has all its elements equal to

\begin{equation}
\label{eq:A8}
    S_{n} = \frac {2}{ L} \int^{\pi}_{0}
            \sin  \left[{\frac {2n{\pi}}{ L}}x({\theta}) \right]
\frac {\sin{\theta}\ d{\theta}}
{\sqrt{{\delta}^{2}+{\gamma}^{2}{\cos}^{2}{\theta}}},
\end{equation}
and $D_{4}=2h \cos{\theta}$ is replaced by $2hC$ where $C$ is 
a symmetric matrix whose n-th codiagonal has all its elements
equal to

\begin{equation}
\label{eq:A9}  
   C_{n} = \frac {2}{ L} \int^{\pi}_{0}  
           \cos  \left[ {\frac {2n{\pi}}{ L}}x({\theta}) \right]
\frac {\cos{\theta}\ d{\theta}}
{\sqrt{{\delta}^{2}+{\gamma}^{2}{\cos}^{2}{\theta}}} .
\end{equation}
An interesting fact is that both $S_{n}$ and $C_{n}$ vanish for
even $n$. The most important terms are those with $n=\pm 1$,
whereas higher-order terms account for distortion of the spiral from its
ideal shape $\theta = {\lambda}x$. Such a distortion occurs even at zero
field in the presence of KSEA anisotropy.

A finite-matrix representation of the differential operator $M$ is then
obtained by restricting the indices $m$ and $n$ to the finite interval
$[-N,N]$ where $N$ may again be chosen as low as 20. The resulting
nonsymmetric $ 4(2N+1)\times 4(2N+1)$ matrix is diagonalized numerically
to yield eigenvalues that are indeed purely imaginary and come in pairs
$\pm i{\omega}$ where ${\omega}={\omega}(q_{1},q_{2})$ is the sought after
physical frequency. We have thus obtained a number of results using
as input the spiral parameters ${\lambda}=1$, ${\gamma}^{2}=1+h^{2}$,
${\delta}= {\delta}(h)$ and $ L= L(h)$ calculated for each field
$h$ as explained in Sec. III. The numerical burden is insignificant and can
be carried out interactively. Explicit results are discussed in Section V.

\section{Vortex states}
\label{sec:Vortex_states}

In the original picture of the CI transition \cite{8} the high-field
commensurate phase is rendered unstable through domain-wall nucleation
at the critical field $h_{c}$ to become a spiral phase for $h < h_{c}$.
The instability at the higher field $h_{2} > h_{c}$ suggested by the
spin-wave analysis of Sec. IV is clearly caused by 2D fluctuations.
Therefore, it is conceivable that the uniform spin-flop phase
is actually destabilized by nucleation of 2D vortices rather than
1D domain walls, as advocated by Bogdanov et al. \cite{21} in a number
of related models.

We thus search for genuinely 2D stationary points of the static
energy that are compatible with $U(1)$ symmetry. First, we introduce
the usual polar coordinates ($r,\psi$) from

\begin{equation}
\label{eq:B1}
 x = \frac {r}{\gamma} \cos{\psi} , \hspace*{1cm}
 y = \frac {r}{\gamma} \sin{\psi} ,
\end{equation}
where the overall rescalling by the constant $\gamma$ will simplify
subsequent calculations. A configuration that is strictly invariant
under the $U(1)$ transformation (\ref{eq:3.4}) reads

\begin{equation}
\label{eq:B2}
  \Theta = {\theta}(r), \hspace*{1cm}
  \Phi   = - \psi ,
\end{equation}
where the minus sign in the second equation is again due
to the peculiar nature of $U(1)$ symmetry in the present problem.
Under normal circumstances, e.g., an isotropic antiferromagnet in an external
field \cite{14}, both choices $\Phi   = \psi$ and $\Phi   = - \psi$ are
compatible with axial symmetry and are referred to as vortex and antivortex.
Here only antivortices are possible within the axially-symmetric ansatz
but will be called vortices for brevity.

When the Ansatz 
(\ref{eq:B2}) is introduced in the potential $V$ of
Eq. (\ref{eq:4.2}) the corresponding total energy $W = \int V dxdy$ reads

\begin{eqnarray}
\label{eq:B3}
W &=& {\pi} \int^{\infty}_{0} r dr
      \left[\ \left( \frac {d{\theta}}{dr} \right)^{2} + 
      \frac {{\sin}^{2}{\theta}}{r^{2}} + {\cos}^{2}{\theta} \right.
     \nonumber
     \\
  & &  - {\nu} \left( \frac {d {\theta}}{dr}
      + \frac { {\cos}{\theta}\ {\sin}{\theta}}{r}\right) \ \Bigg] ,
\end{eqnarray}
where $\nu = 2 {\lambda}/{\gamma}$ is the only relevant parameter
in this static calculation. Also note that we have droped the additive
constant term ${\lambda}^{2}/2$ from the potential (\ref{eq:4.2}) and
thus the energy of the uniform spin-flop state is set equal to zero.
Variation of the energy functional (\ref{eq:B3}) with respect to the
unknown amplitude ${\theta}(r)$ leads to the ordinary differential
equation
\begin{figure}
\centerline{\hbox{\psfig{figure=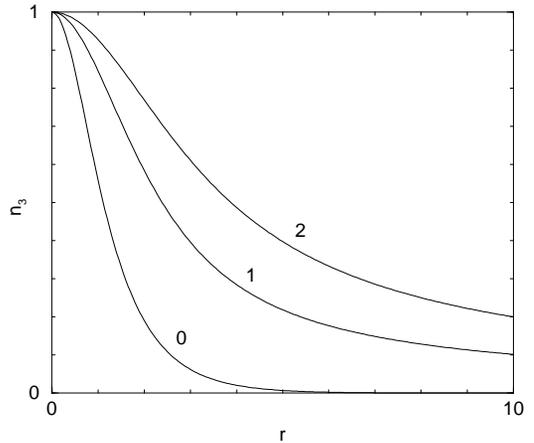,width=7.cm}}}
\vspace*{0.5cm}
\caption{The vortex profile $n_{3} = {\cos}{\theta}$ for three values
 of the parameter ${\nu}=0, 1$ and 2, including the critical value ${\nu}=1$.}
\end{figure}

\begin{equation}
\label{eq:B4}
  r \frac {d^{2}{\theta}}{dr^{2}} + \frac {d{\theta}}{dr}
  + \left(r - \frac{1}{r} \right)\ {\cos}\theta\ {\sin}\theta 
  = {\nu}\ {\sin}^{2}{\theta} ,
\end{equation}
which reduces to the familiar equation for ordinary spin vortices
in the extreme limit ${\nu}=0$. For ${\nu} \neq 0$, solutions
of Eq. (\ref{eq:B4}) exhibit slow decay at large distances, namely
\begin{equation}
\label{eq:B5}
 {\theta}(r) \approx \frac {\pi}{2} - \frac {\nu}{r} + \dots ,
\end{equation}
which turns into exponential decay for ${\nu}=0$. Explicit solutions
were obtained by a straightforward relaxation algorithm and are
illustrated in Fig. 11 for three characteristic values of the parameter
${\nu}$= 0, 1 and 2. 

One may restrict the integral in Eq. (\ref{eq:B3}) to the finite range
$0 < r < R$ and examine its behavior for large $R$. A short calculation
taking into account the asymptotic expansion (\ref{eq:B5}) leads to

\begin{equation}
\label{eq:B6}
 W = {\pi} \left(1 - {\nu}^{2}\right) \ln R + finite\  terms ,
\end{equation}
and thus the energy exhibits the familiar logarithmic divergence.
This asymptotic result demonstrates the crucial role played by
the parameter $\nu$. For ${\nu} < 1$, the energy of a single vortex
is greater than the energy of the uniform spin-flop state by
a logarithmically divergent quantity. This is the usual situation
encountered in the case of ordinary vortices (${\nu} = 0$).
The vortex energy is finite for ${\nu} = 1$ and becomes again
logarithmically divergent but \emph{negative} for ${\nu} > 1$.
The special point ${\nu} = 2 {\lambda}/{\gamma}=1$ leads to the same
critical field $h_{2}$ given earlier in Eq. (\ref{eq:4.6}).

Therefore, for $h < h_{2}$, the energy of the uniform spin-flop
state can be lowered by vortex nucleation. Because of the logarithmic
dependence of the energy on the size of the system, it is clear that
a single vortex cannot by itself produce a thermodynamically significant
effect. Instead, one should expect that a large number of vortices
is created for $h < h_{2}$, probably in the form of a vortex lattice
\cite{21}. We have actually performed several numerical experiments
using the full 2D relaxation algorithm described in the beginning
of Sec. VI. Although we have already obtained some ``spectacular"
pictures indicating the formation of a vortex lattice, we have not
yet been able to lower its energy below that of the nonflat spiral.
It appears that the complete (2D) energy functional displays
glassy behavior in the intermediate region, which may lead to several
nearly degenerate local minima.


\begin{references}
\bibitem[*]{JC_e-mail}Permanent address: Faculty of Science,
 P. J. \v{S}af\'{a}rik University, Park Angelinum 9, 
 040 01 Ko\v{s}ice, Slovakia.
\bibitem[\dagger]{NP_e-mail}Electronic address: papanico@physics.uoc.gr 
\bibitem{1}
 A. Zheludev, G. Shirane, Y. Sasago, N. Kiode, and K. Uchinokura,
 Phys. Rev. B {\bf 54}, 15163 (1996).
\bibitem{2}
 A. Zheludev, S. Maslov, G. Shirane, Y. Sasago, N. Kiode, and K. Uchinokura,
 Phys. Rev. Lett. {\bf 78}, 4857 (1997).
\bibitem{3}
 A. Zheludev, S. Maslov, G. Shirane, Y. Sasago, N. Kiode, K. Uchinokura,
 D. A. Tennant, and S. E. Nagler, Phys. Rev. B {\bf 56}, 14006 (1997).
\bibitem{4}
 A. Zheludev, S. Maslov, G. Shirane, Y. Sasago, N. Kiode, and K. Uchinokura,
 Phys. Rev. B {\bf 57}, 2968 (1998).
\bibitem{5}
 A. Zheludev, S. Maslov, G. Shirane, I. Tsukada, T. Masuda, K. Uchinokura,
 I. Zaliznyak, R. Erwin, and L. P. Regnault,
 Phys. Rev. B {\bf 59}, 11432 (1999).
\bibitem{6}
 I. E. Dzyaloshinskii, Sov. Phys. JETP {\bf 5}, 1259 (1957).
\bibitem{7}
 T. Moriya, Phys. Rev. {\bf 120}, 91 (1960).
\bibitem{8}
 I. E. Dzyaloshinskii, Sov. Phys. JETP {\bf 20}, 665 (1965).
\bibitem{9}
 P. G. de Gennes, Solid State Commun. {\bf 6}, 163 (1968).
\bibitem{10}
 R. B. Meyer, Appl. Phys. Lett. {\bf 14}, 208 (1969).
\bibitem{11}
 P. G. de Gennes and J. Prost, \emph{The Physics of Liquid Crystals}
 (Clarendon Press, Oxford, 1995).
\bibitem{12}
 A. F. Andreev and V. I. Marchenko, Sov. Phys. Usp. {\bf 23}, 21 (1980). 
\bibitem{13}
 V. G. Bar'yakhtar, M. V. Chetkin, B. A. Ivanov, and S. N. Gadetskii,
 \emph{Dynamics of Topological Magnetic Solitons - Experiment and Theory} 
 (Springer Verlag, Berlin, 1994).
\bibitem{14}
 S. Komineas and N. Papanicolaou, Nonlinearity {\bf 11}, 265 (1998).
\bibitem{15}
 J. Chovan and N. Papanicolaou, Eur. Phys. J. B {\bf 17}, 581 (2000).
\bibitem{16}
 T. A. Kaplan, Z. Phys. B {\bf 49}, 313 (1983).
\bibitem{17}
 L. Shekhtman, A. Aharony, O. Entin - Wohlman, 
 Phys. Rev. B {\bf 47}, 174 (1993)
\bibitem{18}
 B. A. Ivanov, A. K. Kolezhuk, and G. M. Wysin,
 Phys. Rev. Lett. {\bf 76}, 511 (1996).
\bibitem{19}
 V. G. Bar'yakhtar and E. D. Stefanovsky, 
 Sov. Phys. Sol. St. {\bf 11}, 1566 (1970).
\bibitem{20}
 A. Bogdanov and A. Shestakov, Low Temp. Phys. {\bf 25}, 76 (1999).
\bibitem{21}
 A. Bogdanov and D. Yablonsky, Sov. Phys. JETP {\bf 69}, 142 (1989);
 A. Bogdanov and A. Hubert, J. Magn. Magn. Mater. {\bf 138}, 255 (1994).
\end{references}
\end{document}